\documentclass[aps,pra,preprint]{revtex4-1}
\usepackage{graphicx}
\usepackage{dcolumn}
\usepackage{bm}
\usepackage{amssymb,amsmath,amsthm,amsfonts}
\usepackage{bbm}
\usepackage{mathtools}
\usepackage{enumitem}
\usepackage{xcolor}
\usepackage{textcomp}
\usepackage{gensymb} 
\usepackage{float}
\usepackage{soul}
\usepackage{booktabs}
\usepackage{natbib}
\usepackage{tikz}
\usetikzlibrary{shapes,arrows,fit}

\begin{document}

%
%
\title{Toward a M{\o}lmer S{\o}rensen 
Gate With .9999 Fidelity}

\author{Reinhold Bl\"umel}
\affiliation{Wesleyan University, Middletown, CT 06459, USA}
\affiliation{IonQ, College Park, MD 20740, USA}

\author{Andrii Maksymov}
\affiliation{IonQ, College Park, MD 20740, USA}

\author{Ming Li}
\affiliation{IonQ, College Park, MD 20740, USA}

\date{\today}

\begin{abstract}
Realistic fault-tolerant quantum computing at reasonable overhead requires two-qubit gates with the highest possible fidelity. Typically, an infidelity of $\lesssim 10^{-4}$ is recommended in the literature. Focusing on the phase-sensitive architecture used in laboratories and by commercial companies to implement quantum computers, we show that even under noise-free, ideal conditions, neglecting the carrier term and linearizing the Lamb-Dicke term in the Hamiltonian used for control-pulse construction for generating M{\o}lmer-S{\o}rensen XX gates based on the Raman scheme are not justified if the goal is an infidelity target of $10^{-4}$. We obtain these results with a gate simulator code that, in addition to the computational space, explicitly takes the most relevant part of the phonon space into account. With the help of a Magnus expansion carried to the third order, keeping terms up to the fourth order in the Lamb-Dicke parameters, we identify the leading sources of coherent errors, which we show can be eliminated by adding a single linear equation to the phase-space closure conditions and subsequently adjusting the amplitude of the control pulse (calibration). This way, we obtain XX gates with infidelities $< 10^{-4}$.
\end{abstract}

\maketitle
 
 

\section{INTRODUCTION}
\label{INTRO}
The trapped-ion architecture, i.e., 
chains of trapped ions, coherently controlled 
via 
Raman hyperfine transitions, 
is one of the most promising 
routes to scalable quantum 
computing 
\cite{NC,UMDQC,BENCHMARKING,AM,ar:LinkePNAS,1-hour,
Honeywell-1,Honeywell-2,Honeywell-3,Honeywell-4}.
This quantum computer architecture is used both 
in laboratory experiments 
and in the 
emerging 
quantum computing 
industry 
\cite{Shantanu,
Honeywell-1,Honeywell-2,Honeywell-3,Honeywell-4}. 
For both the current 
era of 
noisy intermediate-scale 
quantum computing 
\cite{Preskill-1} 
and the anticipated 
era of fault-tolerant, 
error-corrected 
quantum computing 
\cite{NC}, 
two-qubit gates of the 
highest possible 
fidelity are 
essential. 
While 
fault-tolerant 
quantum computing and 
quantum error-correction 
may, in principle, 
be achieved with two-qubit 
gates of 
modest fidelity, the 
overhead, i.e., 
the number of 
physical qubits 
required for one 
error-corrected 
logical qubit depends 
on the native fidelity 
of the physical gates  
and may be enormous 
for physical two-qubit 
gates of only modest 
fidelity. 
A reasonable amount of 
overhead in 
fault-tolerant quantum 
computing can be achieved 
only if the 
physical two-qubit 
gates themselves
have a high native fidelity.  
Typically, for 
realistic, tolerable 
overhead, a physical 
two-qubit infidelity 
of $\lesssim 10^{-4}$ 
is recommended 
\cite{Preskill-1,Knill-1,Monroe-1,Steane-1}.
Two-qubit 
gate infidelities close 
to this target have indeed 
already been 
achieved 
\cite{BEGATE,HIGHF,Brown-1}. 
However, the 
experimental demonstrations 
of high-fidelity two-qubit gates 
are 
restricted to two-qubit gates 
in very short ion 
chains, i.e., chains 
consisting of up to four ions. 
Moreover, to date, even in these 
cases two-qubit gate infidelities 
of $\lesssim 10^{-4}$ have not 
yet been achieved experimentally. 
Two adversaries stand in the way  
of achieving  
two-qubit gate infidelities 
$\lesssim 10^{-4}$: 
Random noise and deterministic, 
coherent 
control errors. 
Even in the shortest chains 
(two to four ions stored 
simultaneously), 
the target of $\lesssim 10^{-4}$ 
infidelity may not be achieved 
if the Hamiltonian used 
to design the control pulses 
for two-qubit gate implementation 
does not accurately enough 
reflect the reality of the 
quantum computer's hardware 
implementation. What this means 
is illustrated in Fig.~\ref{FIG-1}. 
The actual quantum computer, i.e., 
the reality, is governed 
by a Hamiltonian $\hat H_R$. 
Reality can never be captured 
exactly. It can only be modeled 
approximately. Consequently a 
model of the quantum computer is 
constructed (see Section~\ref{HAM}), 
replacing the unknown Hamiltonian 
$\hat H_R$ with $\hat H_M$, where it is hoped 
that $\hat H_M \sim \hat H_R$ 
to a high accuracy. 
Both 
the quantum computer 
($\hat H_R$)
and its model 
($\hat H_M$)
are controlled by 
control pulses that are constructed 
on the basis of a Hamiltonian 
$\hat H_C$. Ideally, 
$\hat H_C=\hat H_R$.  
However, since $\hat H_R$ is unknown,  
the best possible control 
Hamiltonian is 
$\hat H_C=\hat H_M$. However, in most cases 
$\hat H_M$ is too complicated to 
use for 
efficiently 
constructing control pulses 
that frequently also have to be 
computed in real time. 
Therefore, $\hat H_C$ is chosen 
as a compromise, close enough 
to $\hat H_M$ to ensure acceptable control 
of the quantum computer, but simple 
enough to ensure efficient 
control-pulse construction. 
%
\begin{center}
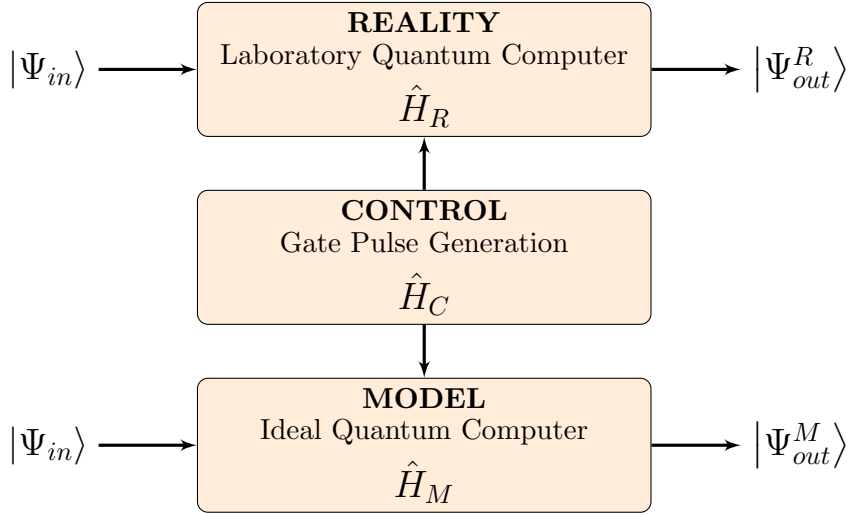
\begin{figure}
\tikzstyle{block} = [rectangle, draw, fill=orange!15,
    text width=15em, text centered, rounded corners, minimum height=4em]
\tikzstyle{line} = [draw, very thick, color=black!100, -latex']

\begin{tikzpicture}[, node distance = 2cm, auto]
    \node [block] (real) {\baselineskip=12pt \textbf{REALITY}\\ Laboratory Quantum Computer\\ \large $\hat{H}_R$};
    \node [left of=real, node distance=5cm] (psirin) {\large $\left| \Psi_{in} \right>$};
    \node [right of=real, node distance=5cm] (psirout) {\large $\left| \Psi_{out}^R \right>$};
    
    \node [block, below of=real, node distance=2.5cm] (cont) {\baselineskip=12pt \textbf{CONTROL}\\ Gate Pulse Generation\\ \large $\hat{H}_C$};
    \node [block, below of=cont, node distance=2.5cm] (model) {\baselineskip=12pt \textbf{MODEL}\\ Ideal Quantum Computer\\ \large $\hat{H}_M$};

    \node [left of=model, node distance=5cm] (psimin) {\large $\left| \Psi_{in} \right>$};
    \node [right of=model, node distance=5cm] (psimout) {\large $\left| \Psi_{out}^M \right>$};
    \path [line] (psirin) -- (real);
    \path [line] (real) -- (psirout);
    \path [line] (cont) -- (real);
    \path [line] (cont) -- (model);
    \path [line] (psimin) -- (model);
    \path [line] (model) -- (psimout);
    
\end{tikzpicture}
\caption {\label{FIG-1} 
Relationship between a hardware-implemented 
quantum computer (REALITY, 
governed by the Hamiltonian $\hat H_R$), 
the corresponding model quantum computer 
(MODEL, governed by 
the model Hamiltonian 
$\hat H_M$) and the control pulses 
(CONTROL, constructed on the basis 
of the control Hamiltonian 
$\hat H_C$) that control both 
the actual hardware-implemented 
quantum computer 
(REALITY) and the 
model quantum computer (MODEL). 
Although controlled by the same 
control pulse, 
since $\hat H_M$ is close to, but 
not equal to $\hat H_R$, a given 
input state $|\psi_{\rm in}\rangle$ 
results in two different output states 
$|\psi^{\rm R}_{\rm out}\rangle 
\neq 
|\psi^{\rm M}_{\rm out}\rangle$, 
which 
depend on whether $|\psi_{\rm in}\rangle$ 
is processed by $\hat H_R$ or 
$\hat H_M$, respectively. 
The overlap 
$|\langle\psi_M|\psi_R\rangle|^2$, 
then, is a measure of how close 
$\hat H_M$ is to $\hat H_R$.} 
\end{figure}
\end{center}
%
%

One of the most basic tasks of 
a quantum computer is to construct 
two-qubit gates 
\cite{NC}. 
In this paper, we focus on 
two-qubit XX gates constructed 
according to 
the M{\o}lmer-S{\o}rensen 
scheme 
\cite{MS-1,MS-2,MS-3}. 
Ideally, given an input state, 
$|\psi_{\rm in}\rangle$, the quantum 
computer, governed by $\hat H_R$, 
is expected to turn 
$|\psi_{\rm in}\rangle$ into 
$|\psi_{\rm out}\rangle=
XX|\psi_{\rm in}\rangle$, 
where XX is the ideal XX gate 
\cite{NC}. 
However, since $\hat H_R\neq \hat H_C$, 
this will not happen in practice. 
Therefore, 
$|\psi^{\rm R}_{\rm out}\rangle$, i.e., 
the output state produced by 
$\hat H_R$,
is different 
from $XX|\psi_{\rm in}\rangle$. 
How different, and given the absence 
of knowledge of $\hat H_R$, can only 
be answered by experimental 
quantum-state tomography 
\cite{QST-1}
and 
is beyond the scope of this paper. 
However, assuming that 
$\hat H_M$ is very close to $\hat H_R$, 
it is possible, at least approximately, 
to assess the quality of the 
control pulses by computing 
the model output state 
$|\psi^{\rm M}_{\rm out}\rangle$ 
and comparing 
it with the ideal output state 
$XX|\psi_{\rm in}\rangle$, 
for instance, 
by computing the overlap 
$|\langle\psi^{\rm M}_{\rm out}|
XX|\psi_{\rm in}\rangle|^2$. 
Only for $\hat H_C=\hat H_M$ do we expect 
this overlap to equal 1. 
However, if $\hat H_M$ is close to 
$\hat H_R$, but $\hat H_C$ is chosen 
as the manageable standard Hamiltonian 
$\hat H_S$
(see Section~\ref{HAM}), we 
expect this overlap and other fidelity measures 
(see Section~\ref{FM}) to differ 
from 1. 
In this case, because of the 
insufficient quality of the 
control pulses, 
the resulting 
two-qubit gates may not be 
accurate enough 
for the target 
infidelity 
(for instance, $< 10^{-4}$). 
To assess this difference, 
and to develop an $\hat H_C$ that 
produces an infidelity 
$\lesssim 10^{-4}$ 
is the purpose of this paper. 
To this end, constructing 
a realistic model Hamiltonian 
$\hat H_M$ 
for Raman-controlled trapped ions, 
we show in this paper 
that currently employed 
control-pulse construction 
techniques based the 
standard Hamiltonian 
$\hat H_S$, 
i.e., $\hat H_C=\hat H_S$ 
(see, e.g., \cite{AMFM})
are not accurate enough to 
achieve the $\lesssim10^{-4}$ infidelity 
target for realistic fault-tolerant 
quantum computing. 
We show this by constructing 
control pulses on the basis of 
$\hat H_S$, which we then use to compute 
two-qubit XX gates in a 7-ion chain 
governed by $\hat H_M$, 
assuming that $\hat H_M$ is 
sufficiently close to 
$\hat H_R$. The XX gate 
simulations of the 7-ion chain are 
performed with a gate-simulator code 
that is accurate on the $10^{-7}$ level 
and takes both computational levels and 
phonon states explicitly into account. 
This way we show that 
even assuming 
ideal conditions, i.e., 
neglecting all incoherent noise 
sources, 
control-pulse construction needs to 
be improved to reach the 
$\lesssim 10^{-4}$ infidelity goal. 

There are two principal methods of implementing 
a trapped-ion chain quantum computer using 
the stimulated Raman scheme: Phase sensitive 
and phase insensitive 
\cite{GREENPAPER}. Both schemes have 
advantages and disadvantages. While 
errors in the phase-insensitive scheme 
were studied before in detail 
\cite{GREENPAPER}, we focus in this paper 
on the phase-sensitive scheme since it 
is technically more straightforward 
to implement and is currently used 
by commercial quantum computing 
companies such as IonQ. Therefore, 
the focus in this paper is to 
isolate the leading coherent error 
sources in the phase-sensitive architecture 
and to construct methods that allow us 
to eliminate these error sources. 
We show in this paper that 
on the basis of 
a new {\it linear} scheme 
of pulse construction, and 
in the  
absence of all incoherent error sources, 
we reach XX gate 
infidelities $\lesssim 10^{-4}$.

Our paper is organized as follows. 
We start 
in Section~\ref{HAM} by presenting 
various Hamiltonians used in gate 
construction and error analysis. 
In Section~\ref{PC} we discuss 
pulse construction methods, 
focusing on 
the AMFM pulse-construction method 
\cite{AMFM}
that we are using throughout 
this paper for generating 
test pulses. 
In Section~\ref{GATESIM} we present 
our gate-simulation method that propagates 
an initial state $|\psi(t=0)\rangle$ 
into a final state $|\psi(\tau)\rangle$, 
where $\tau$ is the gate time, 
including the 2-qubit computational space 
and a portion of the phonon space 
large enough to obtain converged 
results. 
In Section~\ref{FM} we define  
various fidelity measures that we 
use to assess the quality of various 
Hamiltonians used to construct XX gates. 
In Section~\ref{RES} we present our 
numerical simulation results. 
In Section~\ref{CAL} we present a 
simple pulse-scaling method (calibration) 
that may be used to eliminate 
coherent errors that result in XX gates 
that deviate from the target 
degree of entanglement. 
In Section~\ref{MAGEXP} we 
analyze sources of errors that 
are incurred by using a 
pulse function $g(t)$ constructed 
on the basis of the standard Hamiltonian 
to control a quantum computer 
assumed to be governed by 
the full M{\o}lmer-S{\o}rensen 
Hamiltonian. The analysis is 
conducted analytically and 
error integrals are evaluated 
numerically to assess error 
magnitudes. 
In Section~\ref{IMPROVED-CPC} 
we present our {\it linear} method 
of eliminating an important class of 
coherent errors. It is this method, 
together with pulse calibration that 
allows us to suppress coherent XX gate 
errors to the level of $\lesssim 10^{-4}$.
In Section~\ref{SCAL} we investigate 
the scaling of our results to the case 
of chains consisting of 32 ions. 
In Section~\ref{DISC} we 
discuss our results. 
In Section~\ref{CONC} we 
present a brief summary of 
our results and conclude 
the paper. 


\section{HAMILTONIAN}
\label{HAM} 
In this section we derive the 
model Hamiltonian $\hat H_M$, which 
we take to be the full 
M{\o}lmer-S{\o}rensen
Hamiltonian that 
governs a chain of ions in 
a linear Paul trap 
\cite{Shantanu}, 
illuminated 
simultaneously by two laser beams, 
one red detuned and one blue detuned 
\cite{MS-1,MS-2}. 
The derivation starts with 
the effective two-level 
Hamiltonian
\cite{SM-1}
\begin{equation}
\hat H = \hat H_0 + 
\sum_{j=1}^N 
\left(
\frac{\hbar\Omega_{\rm eff}(t)}{2}
\right)
e^{-i[(\Delta {\vec k})\cdot\vec r_j-
\mu t-\Delta\varphi]} 
\hat \sigma_-^{(j)} 
+ 
h.c.,
\end{equation}
obtained via adiabatic 
elimination 
according to 
the Raman $\Lambda$ scheme  
\cite{GREENPAPER}.
Here, 
$N$ is the number of ions 
in the chain, 
\begin{equation}
\hat H_0 = 
    \sum_{p=1}^N 
\hbar\omega_p \hat a_p^{\dagger} 
\hat a_p
\end{equation}
is the phonon Hamiltonian
with the phonon frequencies 
$\omega_p$ and associated 
phonon 
creation 
and destruction operators 
$\hat a_p^{\dagger}$ 
and 
$\hat a_p$, respectively, 
the $j$ sum is 
over the ions in the chain, 
$\Omega_{\rm eff}(t)$ 
is the time-dependent 
effective 
Rabi frequency, 
$\Delta {\vec k}$
is the wavenumber difference 
between the two Raman lasers, 
$\vec r_j$ is the 
position operator 
of ion number $j$ 
in the chain, and 
$\Delta\varphi$ 
is the phase difference between the 
two Raman lasers. 

To construct a  
M{\o}lmer-S{\o}rensen  
(MS) gate \cite{MS-1,MS-2,SM-1}, 
we 
illuminate the ions 
simultaneously with blue 
($+\mu$) and red 
($-\mu$) shifted light 
and, after moving 
to the interaction 
representation with 
respect to $\hat H_0$,
we obtain the MS 
Hamiltonian 
\begin{align}
\hat H_{\rm MS}(t) &= 
\sum_j \hbar\Omega_j 
\cos\left( \mu t - \phi_j^m\right) 
\Bigg\{ 
\cos\left[ 
\sum_p \eta_p^j(\hat a_p^{\dagger} e^{i\omega_p t} + 
\hat a_p e^{-i\omega_p t}) \right] \hat\sigma_y^{(j)} 
\nonumber\\
&+
\sin\left[
\sum_p \eta_p^j(\hat a_p^{\dagger} e^{i\omega_p t} + 
\hat a_p e^{-i\omega_p t}) \right] \hat\sigma_x^{(j)}
\Bigg\} , 
\label{MS-HAM} 
\end{align} 
where $\mu$ is the detuning 
frequency 
\cite{Shantanu}, 
$\eta_p^j$ are the 
Lamb-Dicke parameters 
\cite{Shantanu}, 
and 
$\hat\sigma_x$,
$\hat\sigma_y$,
$\hat\sigma_z$ are 
the Pauli operators.
The Hamiltonian 
(\ref{MS-HAM}) 
may be generalized 
according to 
\begin{equation}
\hbar\Omega_j\cos(\mu t - \phi_j^m) \ \ \ 
\rightarrow \ \ \ 
\hbar\Omega_j(t)\cos\left[\int_0^t 
\mu(t')dt'+\phi_j^{(0)}\right]
\ \ \ 
\rightarrow\ \ \ 
\hbar g_j(t) , 
\label{GENERALIZE} 
\end{equation}
where $g_j(t)$ may 
be any time-dependent 
pulse function, i.e., it 
includes amplitude-modulated 
(AM) pulses
\cite{ROOS-1,AM,AM2}, 
frequency-modulated (FM) pulses 
\cite{FM}, 
phase-modulated (PM) pulses
\cite{PM,PM2,PM3}, 
and simultaneously 
amplitude- and frequency-modulated 
(AMFM) pulses
\cite{AMFM,UMD-PRL}. 
Thus, written in terms 
of the most general 
pulse function $g_j(t)$, 
the MS Hamiltonian 
(\ref{MS-HAM}) 
becomes: 
\begin{equation}
\hat H_{\rm MS} = \sum_j 
\hbar g_j(t) 
\Bigg\{ 
\cos\left[ 
\hat V_j(t)
 \right] \hat\sigma_y^{(j)} + 
\sin\left[
\hat V_j(t) \right] \hat\sigma_x^{(j)}
\Bigg\} , 
\label{HAM-REF}
\end{equation}
where, for later convenience, 
we defined the 
Lamb-Dicke operators 
\begin{equation}
\hat V_j(t) = \sum_p \eta_p^j(\hat a_p^{\dagger} 
e^{i\omega_p t} + 
\hat a_p e^{-i\omega_p t}) ,  
\label{HAM-V}
\end{equation}
which satisfy 
\begin{equation}
\left[ \hat V_j(t_1),\hat V_l(t_2) \right] = 
-2i \sum_p \eta_p^j \eta_p^l 
\sin[\omega_p (t_1-t_2)].  
\label{Vcomm}
\end{equation}
Expanding 
$\hat H_{MS}$ in powers of 
$\hat V$, we obtain a 
family of model 
Hamiltonians 
\begin{equation}
\hat H_M^{(N_c,N_s)} = 
\sum_j \hbar g_j(t) \Bigg\{ 
\sum_{\substack{n=0 \\ n\,{\rm even}}}^{N_c}  
(-1)^{n/2}\,
\frac{\hat\sigma_y^{(j)}} 
{n!} \hat V_j^{n}(t)
+
\sum_{\substack{m=1 \\ m\,{\rm odd}}}^{N_s}  
(-1)^{(m-1)/2}\,
\frac{\hat\sigma_x^{(j)}} 
{m!} \hat V_j^{m}(t)  
\Bigg\} , 
\label{HAM-1}
\end{equation}
where 
\begin{equation}
\hat H_M^{(\infty,\infty)} = \hat H_{MS}. 
\end{equation}
For later convenience, we also 
define the standard Hamiltonian
\begin{equation}
\hat H_S = \hat H_M^{(-2,1)} = 
\sum_j \hbar g_j(t)
\hat V_j(t) \hat\sigma_x^{(j)} , 
\label{HAM-2}
\end{equation}
which is frequently used 
in the literature as the basis for 
control-pulse construction
\cite{ROOS-1,FM,PM,PM2,PM3,AMFM}. 
Here, the superscript 
$N_c=-2$ codes for the 
complete suppression 
(omission) of the 
cosine term in 
(\ref{HAM-1}). 
The test pulses used in this paper 
are also all contructed on the basis 
of $\hat H_S$ (see 
Section~\ref{PC}). 
The expanded model Hamiltonians 
(\ref{HAM-1})
are use in Section~\ref{RES}
to assess the accuracy 
of expansions of $\hat H_{MS}$ 
in powers of the Lamb-Dicke 
operators $\hat V_j(t)$.


\section{CONTROL-PULSE CONSTRUCTION} 
\label{PC} 
To control $\hat H_R$ and $\hat H_M$ 
(see Fig.~\ref{FIG-1}), 
we need a pulse function $g(t)$ 
(see Section~\ref{HAM}). 
Several different 
pulse-construction schemes 
have been proposed in the past, 
including amplitude modulation 
\cite{ROOS-1,AM,AM2}, 
phase modulation \cite{PM,PM2,PM3}, 
frequency modulation 
\cite{FM}, 
and simultaneous 
amplitude- and frequency modulation 
\cite{AMFM,UMD-PRL}. 
All these pulse-construction schemes 
have in common that they are 
based on the 
standard Hamiltonian 
$\hat H_S$ defined in (\ref{HAM-2}). 
For this paper we choose  
sine-AMFM pulses \cite{AMFM,UMD-PRL} 
since they are 
power-optimal and 
straightforward to construct. 
They are defined as 
\begin{equation}
g(t) = \sum_{n_{\rm min}}^{n_{\rm max}} 
B_n \, \sin(\omega_n t), 
\label{PC-1}
\end{equation}
where the basis spans the states 
$n\in\{n_{\rm min},n_{\rm max}\}$, 
$B_n$ are real amplitudes, 
\begin{equation}
\omega_n = \frac{2\pi n}{\tau}  
\end{equation}
are the basis frequencies, 
and 
$\tau$ is the gate time. 
The pulse functions 
(\ref{PC-1}) fulfill 
\begin{equation}
\int_0^{\tau} g(t)\, dt = 0 , 
\label{PC-2}
\end{equation}
an important 
property assumed to hold 
in all of our analytical 
calculations. 

Phase-space closure 
\cite{Shantanu,AMFM,UMD-PRL} 
requires 
\begin{equation}
\alpha_p^j=-\eta_p^j 
\int_0^{\tau} g_j(t) 
\, 
e^{i\omega_p t}\, dt = 0, 
\ \ \ 
p,j=1,\ldots,N. 
\label{PC-3}
\end{equation}
The sine-AMFM pulses (\ref{PC-1}), 
constructed according to 
\cite{AMFM}, 
fulfill (\ref{PC-3}) exactly. 
Throughout this paper we choose 
$N=7$, i.e., we investigate 
the quality of 
two-qubit XX gates in 
a chain of $N=7$ ions. 
$N=7$ was chosen because it 
is large enough to be realistic 
\cite{AMFM,UMD-PRL}, yet small 
enough to enable direct numerical 
simulations of the quantum 
dynamics of the chain 
explicitly including 
a large phonon space. 
In addition, throughout this paper, 
we choose 
$g_j(t)=g(t)$, $j=1,\ldots,N$, 
i.e., whenever we construct 
a two-qubit gate between 
two ions, we assume 
that each of the 
two ions is 
irradiated with 
laser light 
controlled by the 
same pulse function $g(t)$. 
This is a common choice, 
used in laboratories 
\cite{AMFM,UMD-PRL} and 
in commercial 
quantum computers 
\cite{BENCHMARKING}. 

In addition to fulfilling 
the 
phase-space closure 
condition 
(\ref{PC-3}), 
the pulse $g(t)$ 
has to produce the desired degree 
of entanglement $\chi$ 
according to 
\cite{AMFM}
\begin{equation}
\chi=2\sum_{p=1}^N \eta_p^{j_1}\eta_p^{j_2} 
\int_0^{\tau}\, dt \int_0^t\, dt'
g(t) g(t') \sin[\omega_p(t-t')] . 
\label{PC-4}
\end{equation}
A maximally entangling gate 
is obtained for 
$\chi=\pi/4$.
The mode frequencies $\omega_p$, 
$p=1,\ldots,7$, and the 
Lamb-Dicke parameters 
$\eta_p^j$, $p,j=1,\ldots,7$, 
are listed in Tables~\ref{TABLE-omega} 
and \ref{TABLE-eta}, 
respectively. 

For computing analytical infidelities  
we will make 
extensive use of the 
following identities 
that follow 
directly from 
(\ref{PC-3}) 
with the expansion 
(\ref{PC-1}) 
and hold for all 
$p=1,\ldots,N$:
\begin{align}
&\int_0^{\tau} g(t) e^{\pm i\omega_p t}\, dt = 0, 
\label{PC-16}
\\
&\int_0^{\tau} g(t) \sin(\omega_p t)\, dt = 0, 
\\
&\int_0^{\tau} g(t) \cos(\omega_p t)\, dt = 0, 
\\
&\sum_{n_{\rm min}}^{n_{\rm max}} 
B_n \left(
\frac{\omega_n}{\omega_n^2-\omega_p^2}
\right) = 0.
\label{B-sum-identity}
\end{align}
For later use in the following 
sections, we 
also define 
\begin{equation}
\tilde\chi=
\sum_{jp} \left(\eta_p^{j}\right)^2 
\int_0^{\tau}\, dt \int_0^t\, dt'
g(t) g(t') \sin[\omega_p(t-t')] , 
\label{tilde-chi}
\end{equation}
\begin{equation}
G(t) = \int_0^{t}\, g(t') dt' 
=
\sum_n \left(
\frac{B_n}{\omega_n}
\right)
[1-\cos(\omega_n t)] 
= 2
\sum_n \left(
\frac{B_n}{\omega_n}
\right)
\sin^2(\omega_n t/2) , 
\label{big-G-def}
\end{equation}
\begin{equation}
Q(w) = 
\int_0^{\tau} g(t) e^{iwt}\, dt 
= 
\left[ e^{iw\tau}-1\right] 
\sum_n B_n 
\left(
\frac{\omega_n}{w^2-\omega_n^2}
\right) ,  
\label{big-Q-def}
\end{equation}
\begin{align}
f(w) &= 
\int_0^{\tau} g(t) G(t) e^{iwt}\, dt
\nonumber\\
&=\left( \frac{1}{2} \right)
\left[e^{iw\tau}-1\right] 
\sum_{nm}
\left( \frac{B_n B_m }{\omega_n} \right)
\left\{
\frac{\omega_m-\omega_n}
{(\omega_m-\omega_n)^2-w^2}
+
\frac{\omega_m+\omega_n}
{(\omega_m+\omega_n)^2-w^2}
\right\} , 
\label{f-def}
\end{align}
\begin{align}
S_p(w) &= 
\int_0^{\tau} dt_1 
\int_0^{t_1} dt_2 
g(t_1) g(t_2) 
\sin[\omega_p(t_1-t_2)] 
e^{iw t_1}
\nonumber\\ 
&=\int_0^{\tau} dt_1 
\int_0^{t_1} dt_2 
g(t_1) g(t_2) 
\sin[\omega_p(t_1-t_2)] 
e^{iw t_2}
\nonumber\\ 
&=
\sum_{nm}
\frac{B_n B_m}{2(w^2-\omega_m^2)}
\Bigg\{ 
\left(\frac{2\omega_n\omega_m}
{\omega_n^2-4w^2}\right) 
e^{iw\tau} \sin(w\tau) 
\nonumber\\
&+
iw^2
\left(e^{iw\tau}-1\right)
\left[ 
\frac{1}{(\omega_n-\omega_m)^2-w^2}
-
\frac{1}{(\omega_n-\omega_m)^2-w^2}
\right]
\Bigg\} , 
\label{Sp-def}
\end{align}
\begin{equation}
J_p = \int_0^{\tau} 
g(t) G^2(t) e^{i\omega_p t}\, dt , 
\label{Jp-def}
\end{equation}
\begin{align}
Z(w_1,w_2) 
&=  
\int_0^{\tau} dt_1 
\int_0^{t_1} dt_2 
g(t_1) g(t_2) 
e^{iw_1 t_1} e^{iw_2 t_2} 
\nonumber\\
&= 
\sum_{nm} B_n B_m 
\left(\frac{1}
{\omega_m^2 - w_2^2}
\right) 
\Bigg\{ 
\left(\frac{\omega_n\omega_m}
{w_1^2-\omega_n^2}
\right)
\left[ 
e^{iw_1\tau}-1
\right]
\nonumber\\
&-
\left(
\frac{\omega_n\omega_m 
(\omega_m^2-\omega_n^2+w_1^2+ 
4w_1 w_2 + 3w_2^2)}
{\left[ (\omega_n+\omega_m)^2 - 
(w_1+w_2)^2
\right]
\, 
\left[ 
(\omega_n-\omega_m)^2 - 
(w_1+w_2)^2
\right] }
\right)
\, 
\left[ 
e^{i(w_1+w_2)\tau} - 1 
\right]
\Bigg\} , 
\label{Z-def}
\end{align}
and 
\begin{align}
\Phi[\eta,g] &= \sum_{p=1}^N \eta_p^1 
\eta_p^2 \int_0^{\tau} dt_1 
\int_0^{t_1} dt_2 
g(t_1) g(t_2) G(t_2) 
\sin[\omega_p(t_1-t_2)] 
\nonumber\\
&= \frac{\chi\tau}{4\pi}
\sum_{n=n_{\rm min}}^{n_{\rm max}}  
\frac{B_n}{n} , 
\label{Phi-def}
\end{align}
where we used 
(\ref{PC-4}) and 
\begin{equation}
\sum_n \left(
\frac{B_n}{n}
\right) = 
\left(
\frac{2\pi}{\tau^2}
\right)\, 
\int_0^{\tau}dt 
\int_0^t dt'\, g(t')\, 
dt' . 
\label{B-sum-integral-rep}
\end{equation}
%

%
\begin{center}
\begin{table}
\begin{tabular}{|c|c|} 
 \hline
  mode number $p$ & $\omega_p/(2\pi)$ (MHz) \\ 
 \hline 
 1 & 2.953 \\ 
 \hline 
 2 & 2.966 \\ 
  \hline 
 3 & 2.984 \\ 
   \hline 
 4 & 3.006 \\
    \hline 
 5 & 3.029 \\ 
     \hline 
 6 & 3.049 \\ 
 \hline
  7 & 3.060 \\ 
 \hline
\end{tabular}
\caption{Motional-mode frequencies 
$\omega_p/(2\pi)$ 
(in MHz),
$p=1,\ldots,7$,
used 
to construct the 7-ion AMFM test pulses.} 
\label{TABLE-omega}
\end{table}
\end{center}

%
\begin{center}
\begin{table}
\begin{tabular}{ |c||c|c|c|c|c|c|c| } 
 \hline
  $p $ & $j=1$ & $j=2$ & $j=3$ & $j=4$ & $j=5$ & $j=6$ & $j=7$ \\ 
 \hline 
 1 & 0.01033 & -0.03355 & 0.05478 & -0.06313 
   & 0.05478 & -0.03355 & 0.01033 \\
    \hline 
 2 & 0.02226 & -0.05627 & 0.05036 & 0.00000
   & -0.05036 & 0.05627 & -0.02226 \\
 \hline 
 3 & -0.03494 & 0.05644 & 0.00760 & -0.05823  
   & 0.00760 & 0.05644 & -0.03494 \\
  \hline 
  4 & 0.04644 & -0.03074 & -0.05488 & 0.00000  
   & 0.05488 & 0.03074 & -0.04644 \\ 
   \hline 
  5 & 0.05503 & 0.00994 & -0.03678 & -0.05637 
   & -0.03678 & 0.00994 & 0.05503 \\ 
    \hline 
  6 & -0.05848 & -0.04491 & -0.02433 & 0.00000 
   & 0.02433 & 0.04491 & 0.05848 \\ 
     \hline 
  7 & 0.04143 & 0.04143 & 0.04143 & 0.04143 
   & 0.04143 & 0.04143 & 0.04143 \\ 
 \hline
\end{tabular}
\caption{Lamb-Dicke parameters $\eta_p^j$, 
$p,j=1,\ldots,7$,
used 
to construct the 7-ion AMFM test pulses.} 
\label{TABLE-eta}
\end{table}
\end{center}

\section{GATE SIMULATOR} 
\label{GATESIM}
Given a Hamiltonian and 
enough computer power, one 
can always solve the time evolution 
of the computational states including 
phonon excitations to any desired 
accuracy. In this paper we focus on 
a 7-ion chain and expand the complete 
state $|\psi(t)\rangle$
in the combined Hilbert space 
of computational states 
$|a,b\rangle$, $a,b\in \{0,1\}$, 
and phonon 
states $|m_1 m_2 \ldots m_7\rangle$,  
$m_j=0,1,\ldots,m_j^{\rm max}$, 
$j=1,\ldots,7$,
according to 
\begin{equation}
|\psi(t)\rangle = 
\sum_{\substack{a,b \\
m_1, m_2, \ldots, m_7}} 
A^{(a,b)}_{m_1 m_2 \ldots m_7}(t) |a,b\rangle 
 |m_1 m_2 \ldots m_7\rangle ,  
\label{GATESIM-1}
\end{equation}
where $m_j^{\rm max}\geq 0$ is the maximal 
phonon occupation number included in 
the basis. We call 
$\{m_1^{\rm max},\ldots,m_7^{\rm max}\}$ 
the phonon scheme. 
With (\ref{GATESIM-1}), the time-dependent 
Schr\"odinger equation 
results in the 
amplitude equations 
\begin{equation}
i\hbar \dot A^{(a,b)}_{n_1 n_2 \ldots n_7}(t) = 
\sum_{\substack{a',b' \\
m_1, m_2, \ldots, m_7}} 
\langle a,b|\langle n_1\ldots n_7|\hat H(t)|
m_1\ldots m_7\rangle |a',b'\rangle 
A^{(a',b')}_{m_1\ldots m_7}(t) , 
\label{GATESIM-2}
\end{equation}
where $\hat H(t)$ may be any of the model 
Hamiltonians defined in 
Section~\ref{HAM}. 
The set of equations (\ref{GATESIM-2}) 
are ordinary first-order  
differential equations that may 
be solved with any standard 
numerical differential equations solver. 
For simplicity and straightforward 
numerical error control, 
we chose an elementary 
fourth-order 
Runge-Kutta solver with constant 
step size
\cite{NumRec}. 
With this simple 
integrator we are able 
to obtain a relative 
accuracy of the numerical 
solution of better than 
$10^{-7}$. 
Obviously, the gate simulator is 
not limited to 7 ions. 
It scales trivially to any 
number of ions. 
For the purposes of this paper we 
chose 7-ion chains as a compromise 
between a reasonably long ion chain 
and manageable computation times. 
The matrix elements that occur in 
(\ref{GATESIM-2}) are computed 
in Section~\ref{GS-MAT-EL} 
(Appendix~A). 

Convergence in the phonon scheme is 
assessed by using the gate simulator 
to compute 
$|\psi(\tau)\rangle$ 
for the 
Hamiltonian 
$\hat H(t)=\hat H_S(t)
=\hat H^{(-2,1)}(t)$. 
In this case, 
we need to obtain full phase-space 
closure, because the control pulses are 
constructed on the basis of 
the same Hamiltonian that 
governs the time evolution. 
Only if 
enough phonon states are included 
in the basis used by 
the gate-simulator, 
will the phase space be closed 
and thus indicate that the 
phonon-space 
truncation according to the 
phonon scheme guarantees an 
accurate result. 


\section{FIDELITY MEASURES}
\label{FM}
The ideal XX gate 
propagator is 
\begin{equation}
\hat U_{\rm ideal} = 
e^{i\chi\sigma_x^{(1)}\sigma_x^{(2)}} , 
\end{equation}
where $\chi$ is the gate angle. 
For starting state $|\psi_0\rangle$
this produces the ideal final state 
\begin{equation}
|\psi_{\rm ideal}\rangle = 
\hat U_{\rm ideal}
|\psi_0\rangle .
\end{equation}
However, a given gate pulse 
$g(t)$, in general, produces a gate 
propagator of the form 
\begin{equation}
\hat U_{\rm actual} = 
e^{i[\chi\sigma_x^{(1)}\sigma_x^{(2)}+
\lambda \hat E]} , 
\end{equation}
where $\hat E$ is a  
hermitian error operator 
and $\lambda$ 
is its strength.
In this case, and for given 
initial state $|\psi_0\rangle$, 
we define the state fidelity 
according to 
\begin{equation}
F_S = \left|
\langle \psi_{\rm ideal}|
\psi_{\rm actual}\rangle
\right|^2 = 
\left|
\langle \psi_0|
\hat U_{\rm ideal}^{\dagger}
\hat U_{\rm actual} 
|\psi_0\rangle
\right|^2 . 
\label{F_S-def}
\end{equation}
There are two types 
of error operators, i.e., those that 
commute with 
$\sigma_x^{(1)}\sigma_x^{(2)}$ 
and those that do not. 
If only a single, commuting 
error operator $\hat E$ is present, 
the state fidelity, up to 
second order in $\lambda$, is 
\begin{equation}
F_S = \left|
\langle\psi_0| 
e^{-i\chi \sigma_x^{(1)}\sigma_x^{(2)}}
e^{i\chi \sigma_x^{(1)}\sigma_x^{(2)}
+\lambda\hat E} |\psi_0\rangle
\right|^2 = 
\left|
\langle\psi_0| e^{i\lambda \hat E} 
|\psi_0\rangle
\right|^2 
= 
1-\lambda^2 \sigma_{\hat E}^2 , 
\label{FID-EST}
\end{equation}
where 
\begin{equation}
\sigma_{\hat E}^2 = 
\langle\psi_0| \hat E^2 |\psi_0\rangle 
- 
\langle\psi_0| \hat E |\psi_0\rangle^2 . 
\label{FID-EST-1}
\end{equation}
This means that the 
state infidelity 
\begin{equation}
\bar F_S = 1-F_S , 
\label{state-infid}
\end{equation}
up to second order in $\lambda$, is
\begin{equation}
\bar F_S =  
\lambda^2 \sigma_{\hat E}^2 . 
\end{equation}
It is proportional to the square 
of the error strength. 

In case the error operator $\hat E$ 
does not commute with 
$\sigma_x^{(1)}\sigma_x^{(2)}$, 
we may use the 
Baker-Hausdorff-Campbell formula 
to obtain, up to second order 
in $\lambda$,
\begin{align}
F_S &= 
\left|
\langle \psi_0 | 
\hat U_{\rm ideal}^{\dagger}
\hat U_{\rm actual}
|\psi_0\rangle
\right|^2 
= 
\left|
\langle \psi_0 | 
e^{-i\chi\sigma_x^{(1)}\sigma_x^{(2)}} 
e^{i\chi\sigma_x^{(1)}\sigma_x^{(2)}
+ i\lambda\hat E} 
|\psi_0\rangle 
\right|^2 
\nonumber\\
&=
\left|
\langle \psi_0 | 
e^{i\lambda\hat C + i\lambda^2 
\hat D+\ldots}
|\psi_0\rangle
\right|^2
\nonumber\\
&= 
\left|
\langle \psi_0 | 
1 + i\lambda(\hat C+\lambda\hat D)
- \frac{1}{2}\lambda^2 
(\hat C+\lambda\hat D)^2 
+ \ldots 
|\psi_0\rangle
\right|^2
\nonumber\\
&= 
1-\lambda^2 \sigma_{\hat C}^2\, ,  
\label{gen-fid}
\end{align}
where $\hat C$ and $\hat D$ are 
hermitian operators that 
can be expressed as linear 
combinations of nested multi-commutators 
of $\sigma_x^{(1)}\sigma_x^{(2)}$ 
and $\hat E$. 
This implies that in this case, 
too, the infidelity 
$\bar F_S$ is proportional to 
$\lambda^2$. 

Most of the error operators 
$\hat E$
that appear in the present 
context 
do not commute with 
$\sigma_x^{(1)}\sigma_x^{(2)}$. 
However, if $\hat E$ can be 
written in the form 
$\hat E=\hat A\hat\Omega$, 
where $\hat A$ is a 
hermitian error 
operator in the computational 
space and $\hat\Omega$ is a 
hermitian 
error operator in the 
phonon space, and if 
$\hat A$
fulfills the 
anti-commutation relation 
\begin{equation}
\left\{ 
\sigma_x^{(1)}\sigma_x^{(2)}, 
\hat A
\right\} = 
\sigma_x^{(1)}\sigma_x^{(2)} \hat A
+ 
\hat A\sigma_x^{(1)}\sigma_x^{(2)} 
=0 , 
\label{anti}
\end{equation}
we can compute $\bar F_S$ explicitly, 
which also provides an 
explicit expression for the 
operator $\hat C$ 
in (\ref{gen-fid}). 
Examples of error operators 
$\hat A$ that fulfill 
(\ref{anti}) are 
\begin{align}
\hat A \in 
\left\{
\sigma_y^{(1)}, \ 
\sigma_y^{(2)},\  
\sigma_z^{(1)},\ 
\sigma_z^{(2)},\ 
\sigma_x^{(1)}\sigma_y^{(2)} , \ 
\sigma_x^{(1)}\sigma_z^{(2)} , \ 
\sigma_x^{(2)}\sigma_y^{(1)} , \ 
\sigma_x^{(2)}\sigma_z^{(1)}
\right\} . 
\label{A-examples}
\end{align}
Since they act in 
different spaces, we have 
\begin{equation}
[\sigma_x^{(1)}\sigma_x^{(2)},\hat\Omega] 
= 0,\ \ \ 
[\hat A,\hat\Omega] = 0.
\label{AO-commu}
\end{equation}
If 
(\ref{anti}) is 
fulfilled, 
we have 
\begin{align}
\hat U_{\rm actual} &= 
e^{i\chi\sigma_x^{(1)}\sigma_x^{(2)} 
+i\lambda\hat A\hat\Omega}
\nonumber\\
&=
\cos(\hat\varphi) + 
i\left(
\frac{\sin(\hat\varphi)}{\hat\varphi}
\right)
\left[
\chi \sigma_x^{(1)}\sigma_x^{(2)} 
+ \lambda\hat A\hat\Omega
\right] , 
\end{align}
where 
\begin{equation}
\hat\varphi = 
\sqrt{\chi^2 + \lambda^2
\hat A^2\hat\Omega^2} . 
\end{equation}
Then, up to second order 
in $\lambda$, we have 
\begin{align}
F_S = 
\Big|
1 &+ i
\left(
\frac{\lambda}{\chi}
\right)
\cos(\chi)\sin(\chi) 
\langle\psi_0|\hat A\hat\Omega
|\psi_0\rangle
\nonumber\\
&+
\left(
\frac{\lambda}{\chi}
\right)
\sin^2(\chi) 
\langle\psi_0|
\sigma_x^{(1)}\sigma_x^{(2)} 
\hat A\hat\Omega
|\psi_0\rangle
- 
\left(
\frac{\lambda^2}{2\chi^2}
\right)
\sin^2(\chi)
\hat A^2 \hat\Omega^2
\Big|^2 . 
\label{nearly-done}
\end{align}
Now, because of 
(\ref{anti}) and 
(\ref{AO-commu}), 
we have 
\begin{align}
\langle\psi_0| 
\sigma_x^{(1)}\sigma_x^{(2)} \hat A
\hat\Omega
|\psi_0\rangle &= 
-\langle\psi_0| \hat A
\sigma_x^{(1)}\sigma_x^{(2)} 
\hat\Omega
|\psi_0\rangle  
=
-\langle\psi_0|
\hat\Omega
\sigma_x^{(1)}\sigma_x^{(2)} 
 \hat A
|\psi_0\rangle ^* 
\nonumber\\
&=
-\langle\psi_0|
\sigma_x^{(1)}\sigma_x^{(2)} 
 \hat A\hat\Omega
|\psi_0\rangle ^* . 
\end{align}
This means that 
$\langle\psi_0| 
\sigma_x^{(1)}\sigma_x^{(2)} 
\hat A\hat\Omega
|\psi_0\rangle$ 
is purely imaginary, and 
we may write 
\begin{equation}
\langle\psi_0| 
\sigma_x^{(1)}\sigma_x^{(2)} 
\hat A\hat\Omega
|\psi_0\rangle = i \ 
\Im \left\{
\langle\psi_0| 
\sigma_x^{(1)}\sigma_x^{(2)} 
\hat A\hat\Omega
|\psi_0\rangle
\right\} . 
\end{equation}
With this result, we now have, 
up to second order in 
$\lambda$: 
\begin{align}
F_S &= 1 - 
\left(
\frac{\lambda^2}{\chi^2} 
\right)
\sin^2(\chi) 
\langle\psi_0|
\hat A^2\hat\Omega^2 
|\psi_0\rangle
\nonumber\\
&+
\left(
\frac{\lambda^2}{\chi^2} 
\right)
\sin^2(\chi)
\left[ 
\cos(\chi)
\langle\psi_0|\hat A\Omega|\psi_0\rangle
-i \sin(\chi) 
\langle\psi_0|\sigma_x^{(1)}\sigma_x^{(2)}
\hat A\Omega|\psi_0\rangle
\right]^2
\nonumber\\
&= 1-\lambda^2 \sigma_{\hat C}^2 , 
\label{FS-result}
\end{align}
where 
\begin{align}
\hat C &= 
\left(
\frac{\sin(\chi)}{\chi}
\right) 
\left[ 
\cos(\chi)\hat A\hat\Omega - i \sin(\chi) 
\sigma_x^{(1)}\sigma_x^{(2)}
\hat A\hat\Omega 
\right] 
\nonumber\\
&= 
\left(
\frac{\sin(\chi)}{\chi}
\right) 
e^{-i\chi\sigma_x^{(1)}\sigma_x^{(2)}}
\hat A\hat \Omega
. 
\label{expl-C}
\end{align}
Notice that, because of 
(\ref{AO-commu}), 
and since $\hat A$ 
is assumed to fulfill 
(\ref{anti}), 
$i\sigma_x^{(1)}\sigma_x^{(2)}
\hat A\hat\Omega$ 
is a hermitian operator. 
So, $\hat C$ is hermitian, 
as it should be, 
and $\hat C^2=
\hat C^{\dagger}\hat C$. 
With this result, we have 
explicitly 
\begin{equation}
\bar F_S = 
\left(
\frac{\lambda\sin(\chi)}{\chi}
\right)^2 
\left\{
\langle \psi_0|\hat A^2 
\hat \Omega^2|\psi_0\rangle - 
\langle \psi_0| 
e^{-i\chi 
\hat \sigma_x^{(1)}
\hat \sigma_x^{(2)}}
\hat A \hat \Omega 
|\psi_0\rangle^2
\right\} . 
\label{C-squared}
\end{equation}
The explicit form 
(\ref{FS-result})   
of $F_S$ 
confirms the general 
result (\ref{gen-fid}) 
in cases where 
the condition 
(\ref{anti}) is met. 
Many of the most important 
error operators ocurring 
in this paper 
satisfy 
(\ref{anti})
and thus 
(\ref{C-squared})
is applicable. 

In the case of a two-qubit gate, 
the error operator takes the form 
$\hat E=\sum_{j=1,2}
\hat A^{(j)}\hat \Omega^{(j)}$. 
In this case (\ref{FS-result}), 
(\ref{expl-C}), and (\ref{C-squared}) 
may immediately be generalized using 
the substitution 
\begin{equation}
\hat A\hat \Omega \rightarrow 
\sum_{j=1,2}\hat A^{(j)} 
\hat \Omega^{(j)}. 
\end{equation}

The output of the gate simulator code is 
the complete state 
$|\psi^{\rm M}_{\rm out}(\tau)\rangle$, 
which includes 
computational states and phonon states. 
This way, we possess the complete state 
information, which can be used to 
compute the state fidelity 
defined in (\ref{F_S-def}) with 
$|\psi_{\rm actual}\rangle=
|\psi^{\rm M}_{\rm out}(\tau)\rangle$.
In Section~\ref{RES} we use 
the state fidelity $F_S$  
to assess the quality of the various 
model Hamiltonians 
$\hat H_M^{(N_c,N_s)}$.

The state fidelity $F_S$, 
as defined in 
(\ref{F_S-def}), 
depends 
on the initial state $|\psi_0\rangle$.
A more global measue of the 
fidelity of the quantum 
process that implements the 
two-qubit XX gate is the 
process fidelity 
\begin{equation}
F_P = \frac{1}{16} {\rm Tr}
\left[
{\cal E}_{\rm exact}^{\dagger}
{\cal E}_{\rm actual}\right] , 
\end{equation}
where ${\cal E}_{\rm exact}$ is 
the exact XX gate process and 
${\cal E}_{\rm actual}$
is the actual two-qubit XX-gate 
process as computed with the 
gate simulator as described 
in Section~\ref{GATESIM}.

A related fidelity measure, 
also used in Section~\ref{RES}, 
is the 
average gate fidelity $F_G$. 
In our case, 
following \cite{AVGATE}, 
$F_G$ is defined according to 
\begin{equation}
F_G = \frac{1}{80} \sum_{j=1}^{16}
{\rm Tr}
\left[
\hat U_{\rm exact} 
\hat U_j^{\dagger} 
\hat U_{\rm exact}^{\dagger} 
{\cal E}(\hat U_j)
\right], 
\end{equation}
where $\hat U_j$, $j=1,\ldots,16$,
is an operator basis in the 
computational space 
as defined in 
\cite{AVGATE}.

In addition to the 
state infidelity 
$\bar F_S=1-F_S$, defined in 
(\ref{state-infid}), 
we define the infidelities 
\begin{align}
\bar F_G &= 1-F_G, 
\nonumber\\
\bar F_P &= 1-F_P.  
\end{align}
For characterizing the 
quality of the various 
Hamiltonians investigated 
in this paper, it is useful 
to define the 
error in the gate angle 
\begin{equation}
\Delta\chi = \chi-\frac{\pi}{4}  , 
\end{equation}
where $\chi$ is the actual 
gate angle computed by running 
the XX gate simulator 
(see Sections~\ref{GATESIM} 
and \ref{RES}). 
A positive $\Delta\chi$ corresponds to 
an over-rotated gate angle 
$\chi$, while a negative 
$\Delta\chi$ corresponds to an 
under-rotated gate angle $\chi$. 
The error operator associated with 
$\Delta\chi$ is 
$\hat E_{\chi}=\Delta\chi 
\hat \sigma_x^{(1)}
\hat \sigma_x^{(2)}$. 
Then, according to 
(\ref{FID-EST}), 
for $|\psi_0\rangle=
|00\rangle|{\rm ph}\rangle$, for 
instance, where $|00\rangle$ 
is the computational state and 
$|{\rm ph}\rangle$ is any normalized 
phonon state, the state 
infidelity caused by 
$\hat E_{\chi}$ is 
\begin{equation}
\bar F_S^{(\chi)} = (\Delta\chi)^2 . 
\label{chi-infid}
\end{equation}
%


\section{NUMERICAL RESULTS}  
\label{RES} 
As described in 
Section~\ref{PC}, 
we computed 
AMFM pulse functions 
$g(t)$
for gate times ranging from 
$\tau=100\mu$s to $\tau=600\mu$s 
in steps of $100\mu$s, and used 
these pulse functions 
in the XX gate simulator 
(see Section~\ref{GATESIM}). 
As described in 
Section~\ref{GATESIM},  
using the full Hamiltonian 
$\hat H_{MS}$
defined in (\ref{HAM-REF}), 
we 
computed the full state function 
$|\psi(\tau)\rangle$ that includes 
both the computational space 
and the phonon space. 
From $|\psi(\tau)\rangle$, 
as described in 
Section~\ref{FM}, we then 
computed the infidelities 
$\bar F_S$, $\bar F_G$, 
$\bar F_P$, 
and the error $\Delta\chi$ 
in the gate angle $\chi$. 
The results are shown 
in Table \ref{TABLE-FULL}.

%
\begin{center}
\begin{table}
\begin{tabular}{ |c|c|c|c|c|c|c| } 
 \hline
  $\tau$ & 100 & 200 & 300 & 400 & 500 & 600 \\ 
 \hline 
 phonon scheme & 2255111 & 4522111 
 & 2621111
 & 2621111 & 2621111 & 2621111 \\ 
 \hline 
 $\bar F_S$ &  3.1 & 2.6 & 4.0
 & 1.9 & 2.5 & 1.7 \\
  \hline 
 $\bar F_G$ &  2.4 & 2.1 & 3.1
 & 1.5 & 2.0 & 1.3 \\ 
   \hline 
 $\bar F_P$ &  3.0 & 2.6 & 3.8
 & 1.8 & 2.5 & 1.6 \\ 
    \hline 
 $\Delta\chi$ & -0.012 & -0.011 & -0.011  
 & -0.011 & -0.011 & -0.011 \\ 
  \hline
  \hline 
 $\bar F_S^c$ & 1.6       & 1.5       &  2.9
 & 0.78       & 1.4       & 0.61       \\ 
  \hline
  $\bar F_G^c$ & 1.3       & 1.2       & 2.2
 & 0.56       & 1.1       & 0.47        \\ 
  \hline
  $\bar F_P^c$ &  1.6      & 1.5       &  2.7
 & 0.70       & 1.4       &  0.58      \\ 
 \hline
 \hline
  $\bar F_S^{\Phi,c}$ & 0.88       & 0.14       & 0.62
 & 0.41       & 0.34       & 0.45       \\ 
  \hline
  $\bar F_G^{\Phi,c}$ & 0.67       & 0.10       & 0.36
 & 0.28       & 0.26      & 0.34       \\ 
  \hline
  $\bar F_P^{\Phi,c}$ & 0.84       & 0.12       & 0.45
 & 0.34       & 0.32       & 0.42       \\ 
 \hline
\end{tabular}
\caption{Phonon schemes and infidelities 
for XX gate between ion pair (2,5) 
as a function of gate time $\tau$ in 
$\mu$s. 
All infidelities are in units 
of $10^{-4}$. Thus, table 
entries $<1$ correspond to 
infidelities below the 
$10^{-4}$ infidelity target.
$\bar F_S$, $\bar F_G$, 
and $\bar F_P$ are the state-, 
average gate-, and process infidelities 
obtained with control pulses 
constructed on the basis 
of $\hat H_C(t)=\hat H_S(t)$. 
$\Delta\chi$ is negative, 
which indicates that 
$\hat H_C(t)=\hat H_S(t)$ 
under-rotates 
the gate angle $\chi$.
$\bar F_S^c$, $\bar F_G^c$, 
and $\bar F_P^c$ are obtained 
with the calibrated pulse 
$c g(t)$, where $c$ is the 
calibration factor 
[see (\ref{cal-fact})]. 
$\bar F_S^{\Phi,c}$, 
$\bar F_G^{\Phi,c}$, 
and $\bar F_P^{\Phi,c}$ are 
obtained with a 
control pulse 
that is constructed with the 
$\Phi$-condition (\ref{Phi-cond}) 
added and subsequently calibrated. 
The table shows that 
in this case all infidelities 
are suppressed below $10^{-4}$. 
}
\label{TABLE-FULL}
\end{table}
\end{center}
%
Accepting $\hat H_{MS}$ as a good 
approximation of $\hat H_R$,
the most important result we obtain 
from 
Table~\ref{TABLE-FULL} is that 
a gate infidelity $< 10^{-4}$ 
cannot be achieved if the 
control pulses $g(t)$ are constructed 
on the basis of the standard 
Hamiltonian $\hat H_S$. Nevertheless, 
while not quite meeting the goal 
of $\lesssim 10^{-4}$, 
the infidelities obtained are 
very close to this goal. 
We also see that the three different 
infidelity measures, i.e., 
$\bar F_S$, $\bar F_G$, and 
$\bar F_P$ yield similar results 
and any one of them may be used 
as a proxy for assessing the 
infidelity of the two-qubit XX gate. 

Next, we fix the control pulse $g(t)$
(we use the $300\,\mu$s pulse from 
Table~\ref{TABLE-FULL}) and 
determine the quality of the 
various approximations 
$\hat H_M^{(N_c,N_s)}$ to 
the full Hamiltonian 
$\hat H_{MS}$ by computing 
$\bar F_P$ for some of the 
Hamiltonians 
$\hat H_M^{(N_c,N_s)}$. 
The result is shown in 
Table~\ref{TABLE-1}. 
We see that, expectedly, 
$\bar F_P\ll 10^{-4}$ for 
$\hat H_M^{(-2,1)}=\hat H_S$, 
since in this case the pulse 
is generated on the basis of 
$\hat H_S$ and the gate simulator 
is controlled by $\hat H_S$ as well. 
So, ideally, we should obtain 
perfect fidelity. The difference 
from zero in 
Table~\ref{TABLE-1} 
in this case is not due to 
the accuracy of the 
numerical integrator, 
which, as stated in 
Section~\ref{GATESIM} 
is of the order of 
$10^{-7}$, but is due 
to the phonon scheme. 
Including more phonon 
states in our basis 
increases the accuracy 
of our simulations and 
drives the infidelity 
in the case of 
$\hat H_M^{(N_c=-2,N_s=1)}=\hat H_S$
closer to zero. 
The table entry for 
$(N_c=-2,N_s=1)$
also provides us with an estimate of 
the accuracy of the infidelity 
entries in Tables~\ref{TABLE-FULL}
and \ref{TABLE-1}. As indicated 
by the $(N_c=-2,N_s=1)$ entry 
in 
Table~\ref{TABLE-1}, 
the phonon schemes we chose guarantee 
an accuracy of the computed 
infidelities of approximately 
$3\times 10^{-5}$. 

Looking at the infidelity results 
in Table~\ref{TABLE-1} 
for the Hamiltonians expanded 
to zeroth and second orders in 
the $\cos$-term in 
(\ref{HAM-REF}), we see that 
neglecting the zeroth-order term 
in (\ref{HAM-REF}) 
(the carrier term) 
is not justified. But 
we also see 
that, apparently, 
expansion to second order 
of the $\cos$-term 
of (\ref{HAM-REF}) is not 
necessary. This 
is shown analytically 
in more detail 
in Sections~\ref{SECOND}
and \ref{THIRD}.

Turning now to the expansions 
of the $\sin$-term 
of (\ref{HAM-REF}), 
Table~\ref{TABLE-1} 
shows that truncating 
this expansion at the first 
order of the $\sin$ function, 
i.e., linearizing the 
$\sin$-term in the 
Lamb-Dicke parameters, 
is not accurate enough. 
As a consequence, the 
expansion of the 
$\sin$ term in 
(\ref{HAM-REF}) has 
to be carried to 
the third order in 
the Lamb-Dicke parameters, 
but expansion to the 
5th order is not necessary. 

As an overall result of the 
performance tests for different 
$(N_c,N_s)$ model Hamiltonians 
we obtain that 
$\hat H_M^{(0,3)}$
is a good enough approximation 
of $\hat H_{MS}$ on the 
$\lesssim 10^{-4}$ fidelity 
level. Conversely, we also 
obtain the important result 
that pulse construction on 
the basis of the standard 
Hamiltonian 
$\hat H_S=\hat H_M^{(-2,1)}$ 
does not yield infidelities 
$<10^{-4}$. Thus, to improve 
control-pulse construction, 
at a minimum, we have to include 
the carrier term [zeroth-order 
$\cos$ term in (\ref{HAM-REF})]
and the third-order 
$\sin$ term in (\ref{HAM-REF}). 
In Sections~\ref{SECOND} and 
\ref{THIRD}, we 
explore the effects of these 
two additional Hamiltonian terms. 
We also show analytically that 
inclusion of the second- and 
fourth-order terms of the 
Hamiltonian (\ref{HAM-REF}) 
are not needed. 

%
\begin{center}
\begin{table}
\begin{tabular}{ |c|c|c|c|c| } 
 \hline
 $\ $ & $N_s=1$ & $N_s=3$ & $N_s=5$ & $N_s=\infty$ \\ 
 \hline 
 $N_c=-2$ & 0.31 & 1.6 & 1.6 & 1.6 \\ 
 \hline 
 $N_c=0$ & 2.5 & 3.8 & 3.8 & $ - $  \\ 
 \hline 
 $N_c=2$ & 2.5 & 3.8 & 3.8 & $ - $ \\ 
\hline 
 $N_c=\infty$ & $ - $ & $ - $ & $ - $ & $3.8$ \\
 \hline
\end{tabular}
\caption{Process infidelities 
$\bar F_P$ 
in units of 
$10^{-4}$
for phonon scheme 2621111 
and ion pair (2,5) 
produced by model 
Hamiltonians $H_M^{(N_c,N_s)}$. 
A dash in the table means that 
the corresponding quantity was 
not computed. 
} 
\label{TABLE-1}
\end{table}
\end{center}


\section{CALIBRATION}  
\label{CAL} 
Table~\ref{TABLE-FULL} shows 
that 
$\Delta \chi$ is quite large 
and may make a significant contribution 
to the infidelity. 
However, by slightly adjusting 
the amplitude of the control pulse 
$g(t)$, we are able to completely 
eliminate $\Delta\chi$ and thus 
eliminate any contribution to 
the infidelity that otherwise 
would be due to $\Delta\chi$. 
Adjusting $g(t)$ to result 
in $\Delta\chi=0$ is called 
calibration. This procedure, 
extensively used in the laboratory, 
is an attractive way of reducing 
the infidelity, since 
the phase-space closure 
conditions 
(\ref{PC-3}), 
linear in the 
amplitude of the control pulse 
$g(t)$, are invariant 
under a change in the 
control-pulse amplitude. 
So, despite calibration of the 
pulse, phase-space closure 
is always guaranteed exactly, 
independent of the pulse 
amplitude. 

The error operator for 
$\Delta\chi$ is 
$\hat E=\Delta\chi
\hat \sigma_x^{(1)}
\hat \sigma_x^{(2)}$, 
which trivially commutes with 
$\hat \sigma_x^{(1)}
\hat \sigma_x^{(2)}$. 
Therefore, for 
all entries in 
Table~\ref{TABLE-FULL},
the contribution 
of $\Delta\chi$ to 
the infidelity may be 
estimated according to 
(\ref{FID-EST}) 
and 
(\ref{FID-EST-1})
as 
\begin{equation}
\bar F_S^{(\Delta\chi)} = 
(\Delta\chi)^2 
\left[ 
1-\langle\psi_0| 
\sigma_x^{(1)}
\sigma_x^{(2)} 
|\psi_0\rangle^2
\right] .  
\end{equation}
Based on this result, 
and for all pulse lengths 
listed in 
Table~\ref{TABLE-FULL}, 
we have 
$\bar F_S^{(\Delta\chi)}
\sim 1.2\times 10^{-4}$, 
which 
makes a 
significant contribution 
to the infidelities listed 
in 
Table~\ref{TABLE-FULL}. 
However, as mentioned above, 
this infidelity may be 
eliminated by  
calibrating the control 
pulse. Denoting by 
$\chi_{\rm target}$ 
the desired degree of 
entanglement and 
by 
$\chi_g=
\chi_{\rm target} + \Delta\chi$ 
the degree of entanglement 
actually obtained with 
the control pulse $g(t)$, 
the calibration factor 
$c$,
i.e., the (real) factor by which 
the amplitude of $g(t)$ 
has to be multiplied to 
eliminate $\Delta\chi$ 
is obtained 
explicitly, with 
(\ref{PC-4}), as 
\begin{equation}
c = (\chi_{\rm target}/\chi_g)^{1/2} 
= \left(
\frac{\chi_{\rm target}}
{\chi_{\rm target}+\Delta\chi}
\right)^{1/2} . 
\label{cal-fact}
\end{equation}
The entries for 
$F_S^c$, $F_G^c$, and 
$F_P^c$ in 
Table~\ref{TABLE-FULL}
represent the results  
for the respective infidelities 
computed with the same  
control 
pulses used for the 
corresponding entries 
$F_S$, $F_G$, and 
$F_P$, but multiplied 
(calibrated) with the 
calibration factor $c$, 
computed according to 
(\ref{cal-fact}) with 
the corresponding 
$\Delta\chi$ values 
from 
Table~\ref{TABLE-FULL} 
and $\chi_{\rm target}=\pi/4$. 
As expected, 
the result is a 
significant 
reduction of the 
infidelity. 
Since the starting state 
$|\psi_0\rangle$ for the 
results in 
Table~\ref{TABLE-FULL} 
is $|\psi_0\rangle = 
|00\rangle |0\rangle_{\rm ph}$, 
we expect a reduction 
by $\Delta\chi^2$. 
This agrees well with the 
results in 
Table~\ref{TABLE-FULL}. 
Since they can no longer be 
due to $\Delta\chi$, 
the rest of the 
infidelities 
in 
Table~\ref{TABLE-FULL} 
have to come from 
other sources that are not 
proportional to $(\Delta\chi)^2$. 
To look for these sources, 
we now compute the XX-gate 
propagator 
via a Magnus expansion 
as outlined in the following 
section.


\section{MAGNUS EXPANSION} 
\label{MAGEXP}
Given the time-dependent 
Schr\"odinger equation
\begin{equation}
i\hbar \frac{\partial}{\partial t}
|\psi(t)\rangle = 
\hat H(t) |\psi(t)\rangle , 
\label{IMPROVED-1}
\end{equation}
the time evolution 
operator $\hat U(\tau)$ 
of (\ref{IMPROVED-1}) 
over the time interval $\tau$ 
may be constructed systematically 
and analytically, using 
a Magnus expansion
\cite{MAGNUS-PR}, 
i.e., up to 3rd order 
in the Hamiltonian: 
\begin{equation}
\hat U(\tau) = 
\exp\left[ i\hat W_1(\tau) 
+ i\hat W_2(\tau) + 
i \hat W_3(\tau) + 
 \ldots \right] , 
\label{U123}
\end{equation}
where 
\begin{align}
\hat W_1(\tau) &= \frac{1}{i}
\int_0^{\tau} \left(-\frac{i}{\hbar}\right) 
\hat H(t_1)\, dt_1 , 
\nonumber\\
\hat W_2(\tau) &=  
\frac{1}{2i} \int_0^{\tau} dt_1 \int_0^{t_1} dt_2 
\left[ \left(-\frac{i}{\hbar}\right) \hat H(t_1), 
\left(-\frac{i}{\hbar}\right) \hat H(t_2) \right] , 
\nonumber\\ 
\hat W_3(\tau) &= 
\frac{1}{6i} 
\int_0^{\tau}dt_1 \int_0^{t_1}dt_2 
\int_0^{t_2}dt_3
\nonumber\\
&\Bigg\{ \left[ 
\left(-\frac{i}{\hbar}\right) \hat H(t_1),
\left[ \left(-\frac{i}{\hbar}\right) \hat H(t_2), 
\left(-\frac{i}{\hbar}\right) \hat H(t_3) 
\right]\right] 
\nonumber\\
&+\left[ 
\left(-\frac{i}{\hbar}\right) \hat H(t_3),
\left[\left(-\frac{i}{\hbar}\right) \hat H(t_2),
\left(-\frac{i}{\hbar}\right) \hat H(t_1)
\right]\right] 
\Bigg\}   
\label{MAGNUS-U}
\end{align}
are hermitian operators. 
 
In this section, we aim at a 
consistent Magnus expansion up to 
4th order in the Lamb-Dicke parameters 
$\eta_p^j$. Therefore, 
expanding the cosine-term in 
(\ref{HAM-REF}) up to fourth order 
in $\hat V_j$ and the 
sine-term in 
(\ref{HAM-REF}) up to 
third order in 
$\hat V_j$ we use the 
Hamiltonian 
\begin{equation}
\hat H(t) = \hbar g(t) 
\sum_{\alpha=0}^4 
\hat h_{\alpha}(t) 
\label{HAM-h}
\end{equation}
in the Magnus expansion 
(\ref{MAGNUS-U}), 
where
\begin{align}
\hat h_0(t) &=  
 \sum_j \sigma_y^{(j)} , 
\nonumber\\
\hat h_1(t) &= \sum_j \hat V_j(t) 
\hat\sigma_x^{(j)} , 
\nonumber\\
\hat h_2(t) &= 
-\frac{1}{2}\sum_j \hat V^2_j(t) 
\hat\sigma_y^{(j)} , 
\nonumber\\
\hat h_3(t) &= -\frac{1}{6}\sum_j \hat V_j^3(t) 
\hat\sigma_x^{(j)} , 
\nonumber\\
\hat h_4(t) &= \frac{1}{24}\sum_j \hat V_j^4(t) 
\hat\sigma_y^{(j)} . 
\label{OME-DEF}
\end{align}
Notice that $\hat h_0(t)$ is 
actually 
time independent, but we 
formally keep 
the time argument for notational 
convenience. 
Based on the results listed 
in Table~\ref{TABLE-1},
we know that the zeroth order of 
$\cos[\hat V_j(t)]$ contributes 
substantially to the infidelity, 
while the 2nd and higher orders of 
$\cos[\hat V_j(t)]$ 
do not. 
We also know that the 
first and third 
orders of 
$\sin[\hat V_j(t)]$ contribute 
substantially, while the 
5th and higher orders do not.
Thus, the Hamiltonian 
(\ref{HAM-h}) covers all 
these important cases. 
According to Table~\ref{TABLE-1},
it may not be necessary to 
include the second- and 
fourth-order expansion 
terms of the cosine function in 
(\ref{HAM-REF}). However, 
for a consistent expansion 
up to fourth order in the 
Lamb-Dicke parameters,
and also to show  
that these terms are negligible, 
we include them in our expansion 
(\ref{HAM-h}) 
of (\ref{HAM-REF}). 
 
We start in Section~\ref{FIRST} 
by computing $\hat W_1(\tau)$ 
on the basis of (\ref{HAM-h}), 
followed in Sections~\ref{SECOND} 
and \ref{THIRD} by constructing 
$\hat W_2(\tau)$ and $\hat W_3(\tau)$, 
respectively. 
We will find that many of the 
resulting error terms are 
negligible. But we will also 
identify the most significant 
terms that make significant 
contributions to the infidelity. 

With the operators 
defined in (\ref{OME-DEF}), 
we define the hermitian operators 
\begin{equation}
\hat T_{\alpha} = -\int_0^{\tau} 
\hat h_{\alpha}(t)\, g(t)\, dt , 
\ \ \ \alpha=0,1,2,3,4 ,  
\label{T-one-idx}
\end{equation}
\begin{equation}
\hat T_{\alpha\beta} = 
-\frac{1}{2i} 
\int_0^{\tau} dt_1 
\int_0^{t_1} dt_2\, 
g(t_1) g(t_2) 
[\hat h_{\alpha}(t_1),\hat h_{\beta}(t_2)], 
\ \ \ \alpha,\beta=0,1,2,3,4, 
\label{T-second}
\end{equation}
and 
\begin{align}
\hat T_{\alpha\beta\gamma} &= 
\left(\frac{1}{6}\right)
\int_0^{\tau} dt_1 
\int_0^{t_1} dt_2
\int_0^{t_2} dt_3
g(t_1) g(t_2) g(t_3) 
\nonumber\\
&\left\{
[\hat h_{\alpha}(t_1),[\hat h_{\beta}(t_2)],
\hat h_{\gamma}(t_3)] + 
[\hat h_{\alpha}(t_3),[\hat h_{\beta}(t_2)],
\hat h_{\gamma}(t_1)]\right\} , 
\ \ \ \alpha, \beta, \gamma =0,1,2,3,4. 
\label{T-def}
\end{align}
In the following sections, 
these operators are used (i) to 
compute $\hat W_1$, $\hat W_2$, 
and $\hat W_3$ and (ii) 
to compute infidelity 
contributions to the 
gate evolution operator 
$\hat U(\tau)$ defined in 
(\ref{U123}). 


\subsection{First Order}
\label{FIRST}
With the definition in 
(\ref{T-one-idx}) we have 
\begin{equation}
\hat W_1(\tau) =  \sum_{\alpha=0}^4 
\hat T_{\alpha}(\tau). 
\end{equation}
We use 
(\ref{PC-2}) 
together with 
\begin{equation}
\int_0^{\tau} g(t) e^{i\omega_p t}\, dt 
=0\ \ \ 
\Rightarrow
\int_0^{\tau}g(t) \hat V_j(t)\, dt = 0, 
\end{equation}
to arrive at 
\begin{equation}
\hat T_0(\tau) = 0, 
\label{T0}
\end{equation}
\begin{equation}
\hat T_1(\tau) =0, 
\label{T1}
\end{equation}
\begin{equation}
\hat T_2(\tau) = 
\left(\frac{1}{2}\right) 
\sum_{pqj} \eta_p^j \eta_q^j 
\hat\sigma_y^{(j)} 
\left[ Q(\omega_p+\omega_q) 
\hat a_p^{\dagger}\hat a_q^{\dagger} 
\ +\ h.c. \right] 
+ 
\sum_{pqj,p\neq q}
\eta_p^j \eta_q^j 
\hat\sigma_y^{(j)}  
Q(\omega_p-\omega_q) 
\hat a_p^{\dagger}\hat a_q ,
\label{T2}
\end{equation}
\begin{align}
\hat T_3(\tau) &=  
\frac{1}{6}
\sum_{j=1,2}
\hat\sigma_x^{(j)} 
\int_0^{\tau} g(t) \hat V_j^3(t)\, dt  
\nonumber\\
&=
\frac{1}{6} 
\sum_{p,q,r,j} 
\eta_p^j \eta_q^j \eta_r^j 
\hat\sigma_x^{(j)} 
\Bigg\{ 
Q(\omega_p+\omega_q+\omega_r) 
\hat a_p^{\dagger}\hat a_q^{\dagger} 
\hat a_r^{\dagger} 
+
Q(\omega_p+\omega_q-\omega_r) 
\hat a_p^{\dagger}\hat a_q^{\dagger} 
\hat a_r  
\nonumber\\
&+
Q(\omega_p-\omega_q+\omega_r) 
\hat a_p^{\dagger}\hat a_q  
\hat a_r^{\dagger} 
+
Q(-\omega_p+\omega_q+\omega_r) 
\hat a_p \hat a_q^{\dagger} 
\hat a_r^{\dagger} 
\, +\,  h.c.
\Bigg\} , 
\label{T3}
\end{align}
and 
\begin{align}
\hat T_4(\tau) &= 
\left(-\frac{1}{24}\right) 
\sum_j \hat \sigma_y^{(j)} 
\int_0^{\tau} g(t) 
\hat V_j^4(t) \, dt 
\nonumber\\
&= 
\left(-\frac{1}{24}\right) 
\sum_j \hat \sigma_y^{(j)} 
\int_0^{\tau} g(t) \sum_{pqrs} 
\eta_p^j \eta_q^j \eta_r^j \eta_s^j 
\nonumber\\
&\left\{
\hat a_p^{\dagger}\hat a_p^{\dagger}
\hat a_r^{\dagger}\hat a_s^{\dagger} 
e^{i(\omega_p+\omega_q+\omega_r+\omega_s)t} 
+ 
\hat a_p^{\dagger}\hat a_p^{\dagger}
\hat a_r^{\dagger}\hat a_s 
e^{i(\omega_p+\omega_q+\omega_r-\omega_s)t} 
\, 
\ldots
\right\} , 
\label{T4}
\end{align}
where we used $Q(w)$ 
defined in 
(\ref{big-Q-def}), 
in particular, with 
(\ref{PC-2}), 
$Q(0)=0$. 

While $\hat T_0(\tau)$ and 
$\hat T_1(\tau)$ vanish, 
$\hat T_2(\tau)$, 
$\hat T_3(\tau)$, and 
$\hat T_4(\tau)$ need to 
be investigated further since 
they 
may produce undesirable 
contributions to the infidelity 
of the XX gate. 

According to 
(\ref{T2}), 
the size of $\hat T_2(\tau)$ 
is controlled by 
\begin{equation}
\gamma_2 = \max_{pqj,\sigma=\pm 1} 
\left|
\eta_p^j \eta_q^j Q(\omega_p+\sigma\omega_q)
\right| . 
\end{equation}
Numerically, for the 
$300\,\mu$s test pulse, we 
obtain 
\begin{equation}
\gamma_2 = 3.5 \times 10^{-6} . 
\end{equation}
Since $\gamma_2$ is 
significantly  
smaller than $10^{-4}$, 
$\hat T_2(\tau)$ may be neglected. 

According to 
(\ref{T3}), 
the size of 
$\hat T_3(\tau)$ 
is controlled 
by 
\begin{equation}
\gamma_3^{(+)} = 
\left(\frac{1}{6}\right)
\max_{pqrj} 
\left| \eta_p^j \eta_q^j \eta_r^j 
Q(\omega_p+\omega_q+\omega_r)\right| 
\end{equation}
and 
\begin{align}
\gamma_3^{(-)} &= 
\left(\frac{1}{6}\right)
\max_{pqrj} 
\left| \eta_p^j \eta_q^j \eta_r^j 
Q(\omega_p+\omega_q-\omega_r)\right| = 
\left(\frac{1}{6}\right)
\max_{pqrj} 
\left| \eta_p^j \eta_q^j \eta_r^j 
Q(\omega_p-\omega_q+\omega_r)\right|
\nonumber\\
&=
\left(\frac{1}{6}\right)
\max_{pqrj} 
\left| \eta_p^j \eta_q^j \eta_r^j 
Q(-\omega_p+\omega_q+\omega_r)\right| . 
\end{align}
For the $300\,\mu$s test pulse 
we obtain 
\begin{align}
\gamma_3^{(+)} &= 5.8\times 10^{-8}, 
\nonumber\\
\gamma_3^{(-)} &= 3.1\times 10^{-4}. 
\label{gammas}
\end{align}
Since $\hat\sigma_x^{(j)}$ 
commutes with 
$\hat\sigma_x^{(1)}\hat\sigma_x^{(2)}$, 
we may use 
(\ref{FID-EST}) to estimate 
the infidelity 
$\bar F_S^{(T_3)}$ caused 
by $\hat T_3(\tau)$. 
Since, according to 
(\ref{gammas}), both 
$\gamma_3^{(+)}$ 
and $\gamma_3^{(-)}$ 
are smaller than $10^{-3}$, 
and since $\bar F_S^{(T_3)}$, 
according to 
(\ref{FID-EST}), 
involves the square of 
$\hat T_3(\tau)$, 
the infidelity caused by 
$\hat T_3(\tau)$ is 
negligible on the level 
of $10^{-4}$. 

According to (\ref{T4}), the 
size of $\hat T_4(\tau)$ 
is controlled by 
\begin{equation}
\gamma_4 = 
\max_{\substack{j,pqrs \\ 
\sigma_p\sigma_q\sigma_r\sigma_s=\pm 1}}
\left| \eta_p^j \eta_q^j \eta_r^j \eta_s^j 
Q(\sigma_p\omega_p + \sigma_q\omega_q 
+\sigma_r\omega_r + \sigma_s\omega_s)
\right|. 
\end{equation}
For the $300\,\mu$s test pulse we 
obtain 
\begin{equation}
\gamma_4 = 1.1\times 10^{-7}. 
\end{equation}
Therefore, $\hat T_4(\tau)$ 
can be neglected. 

As a result of this 
section we obtain 
that, on the $10^{-4}$ level, 
the first-order terms 
in the Magnus expansion 
do not 
contribute to the 
infidelity. 
We need to be careful however 
to recall that this is only 
true if the pulse function 
$g(t)$ satisfies the condition 
(\ref{PC-2}). Thus, 
(\ref{PC-2}) is an important 
condition that needs to be required for 
high-fidelity XX gates.


\subsection{Second Order}
\label{SECOND}
We now turn to the 
evaluation of the 
second-order terms 
(\ref{T-second}) in the 
Magnus expansion, i.e., we compute 
$\hat W_2(\tau)$ analytically for 
$\hat H(t)$ defined 
in (\ref{HAM-h}) 
up to fourth order in the 
Lamb-Dicke parameters $\eta$.
With the definitions 
(\ref{T-second}) 
of the operators 
$\hat T_{\alpha\beta}(\tau)$, 
the definition 
(\ref{PC-4}) of 
the degree of entanglement, 
making use of 
(\ref{PC-2}), 
and the functions 
defined in 
(\ref{PC-16}) --  
(\ref{B-sum-integral-rep}), 
we obtain 
\begin{equation}
\hat W_2(\tau) = 
\sum_{\alpha,\beta=0}^4 
\hat T_{\alpha\beta}(\tau), 
\end{equation}
where the
operators $\hat T_{\alpha\beta}$, 
listed only 
up to fourth order in 
$\eta$, are  
\begin{equation}
\hat T_{00} = 0,
\label{T00}
\end{equation}
\begin{align}
\hat T_{01} &= 
-\sum_{j=1,2}
\hat\sigma_z^{(j)} 
\int_0^{\tau} g(t) G(t) 
\hat V_j(t)\, dt 
\nonumber\\
&= 
- \sum_{jp} \eta_p^j 
\hat\sigma_z^{(j)} 
\left[ f(\omega_p) \hat a_p^{\dagger}
+ f^*(\omega_p) \hat a_p \right] , 
\label{T01}
\end{align}
\begin{equation}
\hat T_{02} = 0 ,  
\label{T02}
\end{equation}
\begin{align}
\hat T_{03} &= 
\left(\frac{1}{6}\right)
\sum_{j=1,2} 
\hat\sigma_z^{(j)}
\int_0^{\tau} g(t) G(t) 
\hat V_j^3(t)\, dt 
\nonumber\\
&= 
\left(\frac{1}{6}\right)
\sum_{pqrj} 
\hat\sigma_z^{(j)}
\eta_p^j \eta_q^j \eta_r^j 
\Big\{
f(\omega_p+\omega_q+\omega_r) 
\hat a_p^{\dagger}\hat a_q^{\dagger}
\hat a_r^{\dagger} 
+
f(\omega_p+\omega_q-\omega_r) 
\hat a_p^{\dagger}\hat a_q^{\dagger}
\hat a_r  
\nonumber\\
&+
f(\omega_p-\omega_q+\omega_r) 
\hat a_p^{\dagger}\hat a_q 
\hat a_r^{\dagger} 
+
f(-\omega_p+\omega_q+\omega_r) 
\hat a_p\hat a_q^{\dagger}
\hat a_r^{\dagger} 
\ +\ h.c.
\Big\} , 
\label{T03}
\end{align}
\begin{equation}
\hat T_{04} = 0 , 
\label{T04}
\end{equation}
\begin{equation}
\hat T_{10} = \hat T_{01} , 
\label{T10}
\end{equation}
\begin{equation}
\hat T_{11} = 
\tilde\chi \, +\, 
\chi \hat\sigma_x^{(1)}
\hat\sigma_x^{(2)} , 
\label{T11}
\end{equation}
\begin{align}
\hat T_{12} &= 
\left(\frac{1}{4}\right)
\sum_j \hat\sigma_z^{(j)}
\int_0^{\tau} dt_1 
\int_0^{t_1} dt_2 
g(t_1) g(t_2) 
\left\{
\hat V_j(t_1),\hat V_j^2(t_2)
\right\}
\nonumber\\
&- 
\sum_{pq,j\neq k}
\eta_p^j \eta_p^k \eta_q^k 
\hat\sigma_x^{(j)}
\hat\sigma_y^{(k)} 
\left[ 
S_p(\omega_q) \hat a_q^{\dagger}
+
S_p^*(\omega_q) \hat a_q
\right]
,  
\label{T12}
\end{align}
\begin{align}
\hat T_{13} &= 
\left(-\frac{1}{2}\right)
\sum_{jkp} 
\hat\sigma_x^{(j)}
\hat\sigma_x^{(k)}
\eta_p^j \eta_p^k  
\int_0^{\tau} dt_1 
\int_0^{t_1} dt_2\ g(t_1) g(t_2) 
\sin[\omega_p(t_1-t_2)]  
\hat V_j^2(t_1)
\nonumber\\
&=
\left(-\frac{1}{2}\right)
\sum_{jkpqr} 
\hat\sigma_x^{(j)}
\hat\sigma_x^{(k)}
\eta_p^j \eta_p^k 
\eta_q^j \eta_r^j 
[S_p(\omega_q+\omega_r)
\hat a_q^{\dagger}\hat a_r^{\dagger}
\nonumber\\
&+
S_p(\omega_q-\omega_r) 
\hat a_q^{\dagger}\hat a_r + 
S_p^*(\omega_q-\omega_r)
\hat a_q\hat a_r^{\dagger} + 
S_p^*(\omega_q+\omega_r)
\hat a_q\hat a_r 
]  , 
\label{T13}
\end{align}
\begin{equation}
\hat T_{20} = 0,  
\label{T20}
\end{equation}
\begin{equation}
\hat T_{21} = \hat T_{12} , 
\label{T21}
\end{equation}
\begin{equation}
\hat T_{22} = 
\left(\frac{1}{2}\right)
\sum_{jkp} 
\eta_p^j \eta_p^k 
\hat\sigma_y^{(j)}
\hat\sigma_y^{(k)}
\int_0^{\tau} dt_1 
\int_0^{t_1} dt_2
g(t_1)g(t_2) 
\sin[\omega_p(t_1-t_2)]\, 
\left\{
\hat V_j(t_1),\hat V_k(t_2)
\right\} , 
\label{T22}
\end{equation}
\begin{equation}
\hat T_{30} = \hat T_{03} , 
\label{T30}
\end{equation}
\begin{equation}
\hat T_{31} = T_{13} , 
\label{T31}
\end{equation}
and 
\begin{align}
\hat T_{40} = 0. 
\label{T40}
\end{align}

The above explicit, analytical 
results 
are computed making 
explicit use of
(\ref{PC-2}). 

Since $\tilde\chi$ in 
(\ref{T22}) is only a 
c-number, which causes only 
a phase, $\hat T_{11}$ 
generates the desired 
XX gate. All the other 
operators $\hat T_{\alpha\beta}$ 
in 
(\ref{T00}) -- (\ref{T40}), 
if not identically zero, 
are unwanted error operators that generate 
infidelities. 
Because of 
the symmetries 
and the operators 
that are identically zero, 
we have to investigate the 
sizes of only five 
remaining, non-zero error 
operators, i.e., 
$\hat T_{01}$, 
$\hat T_{03}$, 
$\hat T_{12}$, 
$\hat T_{13}$, and 
$\hat T_{22}$.  
 
The size of $\hat T_{01}$ 
is controlled by 
\begin{equation}
\gamma_{01} = \max_{pj} 
\left|\eta_p^j f(\omega_p)\right|. 
\end{equation}
For the $300\,\mu$s test pulse 
we obtain 
\begin{equation}
\gamma_{01} = 1.8\times 10^{-6} . 
\end{equation}
Thus, on the $10^{-4}$ level, 
$\hat T_{01}$ can safely be 
neglected. 

Next, we turn to the 
evaluation of $\hat T_{03}$, which 
proceeds in analogy 
to the evaluation of $\hat T_3(\tau)$
in Section~\ref{FIRST}. 
The size of 
$\hat T_{03}$ is controlled 
by 
\begin{equation}
\gamma_{03}^{(+)} = 
\left(\frac{1}{6}\right)
\max_{pqrj} 
\left| \eta_p^j \eta_q^j \eta_r^j 
f(\omega_p+\omega_q+\omega_r)\right| 
\end{equation}
and 
\begin{align}
\gamma_{03}^{(-)} &= 
\left(\frac{1}{6}\right)
\max_{pqrj} 
\left| \eta_p^j \eta_q^j \eta_r^j 
f(\omega_p+\omega_q-\omega_r)\right| = 
\left(\frac{1}{6}\right)
\max_{pqrj} 
\left| \eta_p^j \eta_q^j \eta_r^j 
f(\omega_p-\omega_q+\omega_r)\right|
\nonumber\\
&=
\left(\frac{1}{6}\right)
\max_{pqrj} 
\left| \eta_p^j \eta_q^j \eta_r^j 
f(-\omega_p+\omega_q+\omega_r)\right| . 
\end{align}
For the $300\,\mu$s test pulse 
we obtain 
\begin{align}
\gamma_{03}^{(+)} &= 7.4\times 10^{-10}, 
\nonumber\\
\gamma_{03}^{(-)} &= 9.3\times 10^{-10}. 
\label{tilde-gammas}
\end{align}
Thus, $\hat T_{03}$ is negligible. 

According to (\ref{T12}), 
the operator $\hat T_{12}$ 
consists of two parts, 
an anti-commutator part and 
a part that originated from 
a commutator. The size of 
the anti-commutator part 
of $\hat T_{12}$
is controlled by 
\begin{equation}
\gamma_{12}^{(a)} = 
\max_{\substack{j,pqr \\ 
\sigma_p\sigma_q\sigma_r=\pm 1}}
\left|
\eta_p^j \eta_q^j \eta_r^j 
Z(\sigma_p\omega_p, 
\sigma_q\omega_q+\sigma_r\omega_r)
\right| . 
\label{gamma12a}
\end{equation}
The commutator part is controlled 
by 
\begin{equation}
\gamma_{12}^{(c)} = 
\max_{pq,j\neq k}
\left|
\eta_p^j \eta_p^k \eta_q^k 
S_p(\omega_q)
\right| . 
\label{gamma12c}
\end{equation}
For the $300\,\mu$s test pulse 
both turned out to be very small. 
Thus, $\hat T_{12}$ can be 
neglected. 

The operator $\hat T_{13}$ 
in (\ref{T13}) can be split 
into a diagonal part 
\begin{align}
\hat T_{13}^{(d)} &=
\left(-\frac{1}{2}\right)
\sum_{jpqr} 
(\eta_p^j)^2  
\eta_q^j \eta_r^j 
[S_p(\omega_q+\omega_r)
\hat a_q^{\dagger}\hat a_r^{\dagger}
\nonumber\\
&+
S_p(\omega_q-\omega_r) 
\hat a_q^{\dagger}\hat a_r + 
S_p^*(\omega_q-\omega_r)
\hat a_q\hat a_r^{\dagger} + 
S_p^*(\omega_q+\omega_r)
\hat a_q\hat a_r 
]   
\label{T13d}
\end{align}
and an off-diagonal part 
\begin{align}
\hat T_{13}^{(o)} &= 
\left(-\frac{1}{2}\right)
\sum_{jkpqr} 
\hat\sigma_x^{(1)}
\hat\sigma_x^{(2)}
\eta_p^{(1)} \eta_p^{(2)} 
[\eta_q^{(1)} \eta_r^{(1)}+ 
\eta_q^{(2)} \eta_r^{(2)}]\, 
[S_p(\omega_q+\omega_r)
\hat a_q^{\dagger}\hat a_r^{\dagger}
\nonumber\\
&+
S_p(\omega_q-\omega_r) 
\hat a_q^{\dagger}\hat a_r + 
S_p^*(\omega_q-\omega_r)
\hat a_q\hat a_r^{\dagger} + 
S_p^*(\omega_q+\omega_r)
\hat a_q\hat a_r 
]  . 
\label{T13o}
\end{align}
The size of $\hat T_{13}^{(d)}$ 
is controlled by 
\begin{equation}
\gamma_{13}^{(d)} = 
\max_{\substack{j,pqr \\ 
\sigma=\pm 1}}
\left|
(\eta_p^j)^2 \eta_q^j \eta_r^j 
S_p(\omega_q+\sigma\omega_r)
\right| . 
\end{equation}
For the $300\,\mu$s test pulse 
we obtain 
\begin{equation}
\gamma_{13}^{(d)} = 1.17\times 10^{-3} . 
\end{equation}
Thus, 
compared to the 
target infidelity 
of 
$\ll 10^{-4}$, 
we cannot dismiss 
$\hat T_{13}^{(d)}$ 
outright. However, 
taking $\hat T_{13}^{(d)}$  
as the error operator and 
since $\hat T_{13}^{(d)}$ 
commutes with 
$\sigma_x^{(1)}\sigma_x^{(2)}$, 
we can use the 
infidelity 
estimate 
(\ref{FID-EST}), 
which involves the 
squares of 
$\hat T_{13}^{(d)}$, 
i.e., the contribution of 
$\hat T_{13}^{(d)}$ to the 
infidelity of $\hat U(\tau)$ 
is expected to be 
of the order of 
$(\gamma_{13}^{(d)})^2\sim 
10^{-6}$. 
This means that the contribution 
of $\hat T_{13}^{(d)}$ 
to the infidelity of 
$\hat U(\tau)$ can be 
neglected. 

We now turn to 
$\hat T_{13}^{(o)}$. 
It contains 
$\sigma_x^{(1)}\sigma_x^{(2)}$ 
and thus contributes to 
over/under rotation of 
the XX gate, i.e., 
it contributes to 
$\Delta\chi$ 
(see Table~\ref{TABLE-FULL}). 
The operator 
$\hat T_{13}^{(o)}$ 
acts in the 
computational space 
but also produces phonon 
excitations. 
However, according to 
(\ref{T13o}), 
the two-phonon exitation terms 
are proportional to 
$S_p(\omega_q+\omega_r)$, 
which are nonresonant and 
very small. 
Assuming that the initial 
state starts out in the 
phonon ground state, 
$|0\rangle_{\rm ph}$, 
the operator 
$\hat T_{13}^{(o)}$ is, 
to an excellent approximation, 
\begin{equation}
\hat T_{13}^{(o)} = 
\left(-\frac{1}{2} \right)
\sum_{pq}
\eta_p^{(1)}\eta_p^{(2)} 
\left[ 
(\eta_q^{(1)})^2 + 
\eta_q^{(2)})^2 
\right] 
\sigma_x^{(1)}\sigma_x^{(2)} 
S_p(0). 
\end{equation}
Now, with 
(\ref{PC-4}), 
\begin{equation}
\sum_p \eta_p^{(1)}\eta_p^{(2)} 
S_p(0) = 
\sum_p \eta_p^{(1)}\eta_p^{(2)} 
\int_0^{\tau}dt_1 
\int_0^{t_1} dt_2 g(t_1)g(t_2) 
\sin[\omega_p(t_1-t_2)] 
= \frac{\chi}{2} . 
\end{equation}
With this result, 
\begin{equation}
\hat T_{13}^{(o)} = 
\left[ 
\left(-\frac{\chi}{4}\right) 
\sum_{jp}
(\eta_p^j)^2
\right] 
\sigma_x^{(1)}\sigma_x^{(2)} . 
\end{equation}
The prefactor of the 
$\sigma_x^{(1)}\sigma_x^{(2)}$ 
operator is a c-number. 
Therefore, 
$\hat T_{13}^{(o)}$ 
produces a contribution to 
the degree of entanglement 
$\chi$. Since we have 
$\hat T_{31}=\hat T_{13}$, 
the total contribution 
to the degree of entanglement 
is 
\begin{equation}
\Delta\chi =  
\left(-\frac{\chi}{2}\right) 
\sum_{jp} (\eta_p^j)^2 . 
\label{Dchi-eta}
\end{equation}
Numerically, for the 
$N=7$ case 
(see Table~\ref{TABLE-eta}), 
we have 
\begin{equation}
\sum_{jp}
(\eta_p^j)^2 = 2.45\times 10^{-2} . 
\end{equation}
For $\chi=\pi/4$, this results in 
\begin{equation}
\Delta\chi = -9.62\times 10^{-3}. 
\label{delt-chi}
\end{equation}
According to our numerical simulations 
(see Table~\ref{TABLE-FULL}), 
we have 
$\Delta\chi \approx -0.011$. 
So, our analytical calculations 
predict the correct sign 
(under-rotation) of $\chi$. 
In addition, the magnitude of 
the relative error of 
our analytical prediction is 
$(0.011-9.62\times 10^{-3})/0.011 
\approx 0.13$, i.e., our 
analytical prediction is only 
of the order of 10\% off. 
 
According to 
(\ref{Dchi-eta}), 
$\Delta\chi$ depends 
only on the Lamb-Dicke parameters, 
and not on the gate duration 
$\tau$. This is reflected in 
Table~\ref{TABLE-FULL} and 
explains why $\Delta\chi$
in Table~\ref{TABLE-FULL}
is approximately constant, 
independent of $\tau$. 

Summarizing the results obtained 
in this Section, we find that 
the second-order Magnus-expansion 
operators yield only two substantial 
contributions, i.e., the operator 
$\hat T_{11}$, which generates the 
desired XX gate and the operator 
$2\hat T_{13}$ which 
explains 
the under-rotation of 
the gate angle 
and, approximately, 
its size, as listed in 
Table~\ref{TABLE-FULL}. 

At the end of Section~\ref{CAL}, 
we noticed that $\Delta\chi$ 
explains a significant contribution 
to the infidelity of 
$\hat U(\tau)$, but cannot 
explain the entire infidelity 
contribution. We took this 
as the 
motivation to look for the 
additional sources of infidelity 
in the various orders of the 
Magnus expansion. 
In this section we found that 
the second order explains only 
the origin of $\Delta\chi$, 
but does not reveal 
any 
additional significant 
sources of infidelity. 
Since the second 
order of the Magnus 
expansion 
did not reveal these sources, 
we now investigate the 
third order of the 
Magnus expansion, which, 
indeed, reveals 
the remaining significant 
sources of infidelity. 
In addition, we will find 
that this 
source of 
infidelity can be eliminated 
by the addition of a 
single linear equation 
to the control-pulse construction 
protocol. 
%

%
\subsection{Third Order}
\label{THIRD}
In this section we compute 
$\hat W_3(\tau)$.  
With the definitions 
(\ref{T-def}) 
stated at the 
end of the introduction 
to Section~\ref{MAGEXP}, 
we have 
\begin{equation}
\hat W_3(\tau) = 
\sum_{\alpha\beta\gamma=0}^4
\hat T_{\alpha\beta\gamma} .
\end{equation}

All operators $\hat T_{jkl}$ 
up to fourth order in $\eta$ 
are listed in Section~\ref{3COMM} 
(Appendix~B). 
There is only a single operator 
of zeroth order in $\eta$, i.e., 
$\hat T_{000}\sim \eta^0$. 
It is 
trivially zero and does not 
contribute to the infidelity. 
 
There are three operators 
of first order in $\eta$, 
i.e., 
$\hat T_{001}$, 
$\hat T_{010}$, 
$\hat T_{100}$, where 
[see Section~\ref{3COMM} 
(Appendix~B)] 
$\hat T_{010}=2\hat T_{001}$, 
and 
$\hat T_{100}=0$. This means 
that only $\hat T_{001}$'s 
contribution  
to the infidelity 
has to be investigated.  
With (\ref{Jp-def}), the 
infidelity caused by 
$\hat T_{100}$ is 
controlled by 
\begin{equation}
\gamma_{001} = 
\max_{jp} 
\left|
\eta_p^j J_p \right| . 
\end{equation}
For the $300\,\mu$s 
control pulse we obtain 
\begin{equation}
\gamma_{001} = 1.2\times 10^{-6}. 
\end{equation}
Therefore, $\hat T_{001}$, 
and with it $\hat T_{010}$, 
are negligible. 
Since $\hat T_{100}=0$, 
all operators $\sim \eta^1$ 
can be neglected. 

There are six operators 
of second order in $\eta$, 
i.e., 
$\hat T_{002}$,
$\hat T_{011}$,
$\hat T_{020}$,
$\hat T_{101}$,
$\hat T_{110}$, and 
$\hat T_{200}$. 
Of these, according to 
Section~\ref{3COMM},  
only 
$\hat T_{011}$, 
$\hat T_{101}$, and 
$\hat T_{110}$ 
are nonzero, and 
of those, only 
$\hat T_{011}$ and 
$\hat T_{110}$ 
are non-negligible. 
While $\hat T_{011}$ 
acts only on the computational 
space, 
$\hat T_{110}$ 
also has a part that 
produces phonon excitations. 
This part, however, is 
negligibly small. 
As an overall result of 
the operators $\sim \eta^2$, 
we obtain that the effective 
error operator corresponding 
to the leading parts of 
$\hat T_{011}$ and 
$\hat T_{110}$ is 
\begin{equation}
\hat E^{\sigma_x\sigma_z} = 
4\Phi[\eta,g]\ 
[\hat\sigma_x^{(1)}\hat\sigma_z^{(2)}
+\hat\sigma_x^{(2)}\hat\sigma_z^{(1)}].
\label{Esxsz}
\end{equation}
We checked that 
all operators of order three 
and four in $\eta$ are negligible. 
Thus, 
(\ref{Esxsz}) is the only 
significant error operator 
that results from the third-order 
Magnus expansion. We note that 
this operator acts only 
in the computational space, 
not in the phonon space. 
With 
(\ref{C-squared}) 
and 
$\lambda=4\Phi[\eta,g]$,
we obtain for the infidelity 
contribution of 
(\ref{Esxsz}) for 
$|\psi_0\rangle=
|00\rangle|0\rangle_{\rm ph}$: 
\begin{equation}
\bar F_S^{(\sigma_x\sigma_z)} = 
\left(
\frac{\lambda\sin(\chi)} 
{\chi} \right)^2 
\langle\psi_0 | 
\left[ 
\sigma_x^{(1)}\sigma_z^{(2)} + 
\sigma_x^{(2)}\sigma_z^{(1)}
\right]^2 | \psi_0\rangle 
= \left(\frac{4\lambda}{\pi}\right)^2 .
\label{bar-Fs-sig-sig}
\end{equation}
For the $300\mu$s test pulse we have 
\begin{equation}
\lambda = 1.17\times 10^{-2} . 
\end{equation}
Thus, with 
(\ref{bar-Fs-sig-sig}) 
we obtain: 
\begin{equation}
\bar F_S^{(\sigma_x\sigma_z)} = 
2.2\times 10^{-4} . 
\end{equation}
Together with 
(\ref{delt-chi}), we 
now obtain an  
estimate for the 
total infidelity 
of the $300\,\mu$s 
test pulse according to 
\begin{equation}
\bar F_S = 
(\Delta\chi)^2 + 
\bar F_S^{(\sigma_x\sigma_z)} 
= (-9.62\times 10^{-3})^2 + 
2.2\times 10^{-4}
= 3.1\times 10^{-4} . 
\end{equation}
According to 
Table~\ref{TABLE-FULL}, 
this accounts for 
about 80\% of the infidelity 
of the $300\,\mu$s test pulse. 
Since, via calibration, 
$\Delta\chi$ can always 
be set to zero, 
generating control pulses 
that zero out the $\Phi$ functional 
may go a long way to reduce 
the infidelity. 
How to generate such control pulses, 
and that this recipe actually 
works to suppress the 
the 
infidelity below $10^{-4}$, 
is shown in the following 
section.


\section{IMPROVED CONTROL-PULSE CONSTRUCTION}
\label{IMPROVED-CPC}
In Section~\ref{CAL} we argued, and 
proved numerically, that a significant 
portion of the infidelity can be 
removed by calibration of the 
control pulse. Here we show 
that by eliminating the infidelity 
due to the $\Phi$-functional 
(see Section~\ref{THIRD}), 
we can further 
suppress the infidelity below 
the level of 
$10^{-4}$. 
To eliminate $\Phi$, we 
add the single linear equation 
\begin{equation} 
  \sum_n \frac{B_n}{n} = 0 
\label{Phi-cond}
\end{equation}
to the AMFM pulse-solver code 
and obtain 
new pulses $\tilde g(t)$ that 
zero out 
$\Phi[\eta,\tilde g]$, defined in 
(\ref{Phi-def}). That 
the AMFM code with 
the condition 
(\ref{Phi-cond}) added 
produces control 
pulses $\tilde g(t)$ with 
$\Phi[\eta,\tilde g]=0$ 
was confirmed explicitly. 
As discribed in Section~\ref{CAL}, 
$\tilde g(t)$ may be renormalized 
(calibrated) such that 
$\tilde g(t)$ not only produces 
$\Phi[\eta,\tilde g]=0$, but 
simultaneously produces 
$\Delta\chi=0$. 
Running our gate-simulator code 
(see Section~\ref{GATESIM}) 
with the calibrated pulses 
$\tilde g(t)$, 
we obtain the infidelities 
$\bar F_S^{\Phi,c}$, 
$\bar F_G^{\Phi,c}$, and 
$\bar F_{\Phi}^{\Phi,c}$ 
as shown in Table~\ref{TABLE-FULL}. 
We see that in all cases the 
calibrated pulses $\tilde g(t)$
produce infidelities 
below $10^{-4}$. 
We note that the calibrated 
pulses $\tilde g(t)$
require only insignificantly larger 
power compared with the 
original pulses $g(t)$. 
This is as expected, since 
only one additional condition, 
i.e., the condition (\ref{Phi-cond}), 
was added 
to the original set of linear 
phase-space closure conditions 
(\ref{PC-3}). 
We point out that, in conjunction 
with calibration, the construction 
of the improved AMFM pulses is still 
a {\it linear} process. 
Nonlinear optimizer codes are 
not required. 


\section{SCALING}
\label{SCAL}
\begin{figure}
\includegraphics[scale=0.8,angle=0]{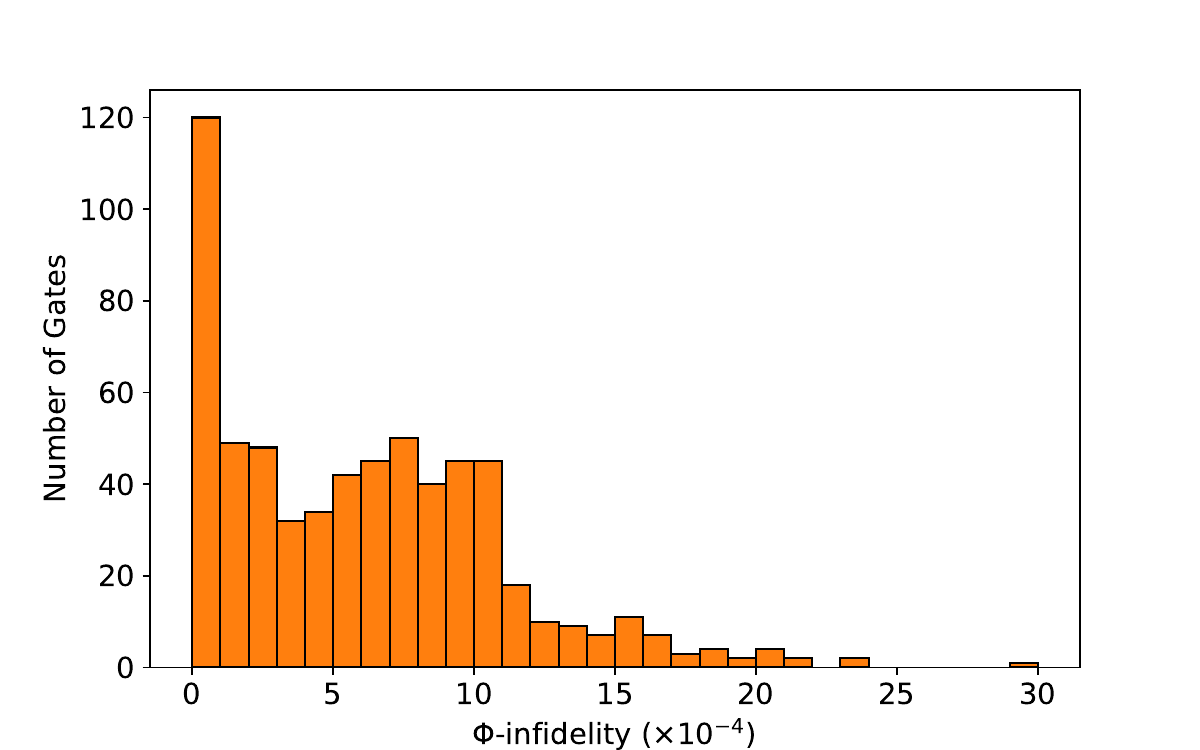}
\caption {\label{FIG-HISTO} 
Number of gates vs. $\Phi$-infidelity for 
$N=36$-ions, uncalibrated, 
$\tau=300\,\mu$s 
AMFM pulses, not requiring 
the $\Phi$ condition 
(\ref{Phi-cond}). The infidelities 
are in units of $10^{-4}$.
The histogram plot illustrates that 
given a $10^{-4}$ fidelity target, 
the infidelities introduced by 
control pulses not requiring 
(\ref{Phi-cond}) are severe for 
a majority of $N=36$ gates. 
Conversely, requiring 
(\ref{Phi-cond}), a single, 
linear equation added to the 
pulse-construction protocol, eliminates 
this type of infidelity completely. 
      }
\end{figure}

So far we have focused entirely on 
the 7-ion case for which we have 
a complete analysis machinery 
in place 
consisting of pulse construction, 
analytical formulas for error estimates, 
and a gate simulator that includes all 
the relevant phonon states 
(see Section~\ref{GATESIM}). 
But how do the control errors scale 
with the number of ions $N$? 
Since the required pulse power increases 
with increasing $N$ 
\cite{AMFM}
but at the same time 
the Lamb-Dicke parameters 
$\eta$ decrease, this is an 
open question. We partially 
answer this question in the 
following way. 
Our analytical results do not 
depend on the number of ions $N$ in 
the chain, i.e., our analytical 
results are valid for any number 
of qubits. In particular, as soon 
as a control pulse is generated 
(it does not matter whether this 
is an AM pulse, FM pulse, AMFM 
pulse, or any other type of pulse), 
this pulse can immediately be inserted 
into our analytical formulas, which then 
may be used to obtain infidelity 
estimates for this particular  
$N$-ion control pulse. 
We illustrate this method 
for $N=36$ by 
computing the infidelity 
(\ref{bar-Fs-sig-sig}), i.e., 
the leading source of 
infidelity,  
for $N=36$ uncalibrated 
control pulses for 2-qubit 
XX gates between all 
possible gate combinations 
$(i_0,j_0)$, 
$j_0=1,\ldots,i_0-1$, 
$i_0=2,\ldots,36$. 
This results in 630 gate 
combinations. The infidelities 
obtained are displayed 
in Fig.~\ref{FIG-HISTO} in the form of 
a bar graph, where the height 
of a bar shows the 
frequency of occurrence of 
infidelities within the width 
of the bar. The infidelities 
in Fig.~\ref{FIG-HISTO}
are in units of 
$10^{-4}=1\,$pptt, 
where the unit ``pptt'' 
denotes one part per ten thousand. 
We see that the infidelities 
generated by $N=36$ control pulses 
constructed on the basis of 
the Standard Hamiltonian 
$\hat H_S$ are 
significant, and in many cases 
they are 
much larger than $10\,$pptt. 
This particular error source, 
in conjunction with calibration, 
can now be eliminated completely 
by using the linear construction 
technique outlined in 
Section~\ref{IMPROVED-CPC}. 
%
The histogram in 
Fig.~\ref{FIG-HISTO}
was made 
for $300\,\mu$s AMFM pulses. 
The question is: How does this 
scale with the gate time $\tau$. 
To answer this question, we also computed 
36-ion AMFM histograms at $700\,\mu$s.  
The result is that most of the 
$\Phi$-infidelities 
for the $700\,\mu$s AMFM histogram are below 
$5\,$pptt. 
About half of the gates 
are good gates with infidelities less than 
$1\,$pptt
and most of the other gates have 
$\Phi$-infidelities less than 
$5\,$pptt. This indicates that 
the $\Phi$-infidelity is a sensitive 
function of gate time $\tau$. 


\section{DISCUSSION}
\label{DISC}
In this paper  
we consider neither stochastic 
nor systematic errors in the 
quantum computer hardware. 
Instead, we ask a different question: 
Even in the absence of all stochastic 
and systematic hardware errors, 
i.e., even in the ideal situation 
that the quantum computer is governed 
exactly by the model Hamiltonian 
$\hat H_M$ (see Fig.~{FIG-1} and discussion 
in Section~\ref{INTRO}), and given 
that the pulse construction is 
based on the Standard Hamiltonian 
$\hat H_S$,
is it even in principle 
possible in this idealized case to 
reach XX gate infidelities 
better than $10^{-4}$ for 
all gates? 
The answer given in this paper 
(see, in particular, Table~\ref{TABLE-FULL} 
and Fig.~\ref{FIG-HISTO}), 
is no. 
However, we also show that 
by slightly modifying 
the pulse construction protocol 
by including only the single additional 
linear condition (\ref{Phi-cond}), and 
subsequently calibrating the pulse obtained, 
it is possible to eliminate the two leading 
sources of infidelity and, at least in 
the 7-ion case considered in detail in this 
paper, then achieve infidelities 
smaller than $10^{-4}$ in all cases 
considered. While, in Section~\ref{SCAL}, 
we demonstrated that the coherent errors 
produced by the error operator 
(\ref{Esxsz}) persist even in the 
$N=36$ case, due to limited computer 
resources, we are not currently able to 
show that, following the new 
pulse-construction protocol, 
the infidelity can be suppressed 
below $10^{-4}$ in this case 
as well. 
However, as an application of our 
analytical formulas stated in 
Sections~\ref{FIRST}, \ref{SECOND}, 
\ref{THIRD}, and \ref{3COMM} (Appendix~B), 
we are confident that even in the 
$N=36$ case 
our two-stage 
protocol of pulse construction, i.e., 
implementing (\ref{Phi-cond}) with 
subsequent calibration (see Section~\ref{CAL}), 
will result in pulses that produce infidelities 
smaller than $10^{-4}$.

 
\section{SUMMARY AND CONCLUSIONS}
\label{CONC} 
In this paper, 
focusing 
on the 
phase-sensitive geometry 
\cite{GREENPAPER}, 
we found that even in 
the absence 
of all experimental 
coherent and incoherent 
errors, the 
control pulses $g(t)$ 
computed on the basis 
of the Standard Hamiltonian 
$\hat H_S$
are not accurate enough 
to consistently generate XX gates 
with infidelities 
$\lesssim 10^{-4}$. 
For ion chains consisting 
of $N=7$ ions (7-qubit case), 
we based this conclusion 
on numerical simulations 
including all relevant 
phonon states, and 
on analytical evaluations 
of error terms generated 
by a third-order Magnus expansion, 
keeping all commutator terms 
up to fourth order in 
the Lamb-Dicke parameters 
$\eta$. 
Not satisfied with this negative 
result, identifying the two 
leading sources of coherent errors 
we defined a new 
control-pulse construction 
protocol obtained by adding 
a single, 
{\it linear} equation to the 
standard phase-space closure 
equations that eliminates both 
$\Delta\chi$ and $\Phi$ errors. 
In all $N=7$-cases studied in 
this paper, pulses generated 
with the new pulse-construction 
protocol produced 
XX-gate infidelities 
$\lesssim 10^{-4}$.  
We also showed that increasing 
the number of ions in the chain 
to $N=36$, the two principal sources 
of coherent control errors remain. 
While our computational resources 
are not currently sufficient to 
run our gate simulator code for the 
$N=36$ case, we are confident that 
even in this case, our 
two-stage method of zeroing out 
the $\Phi$ functional 
(\ref{Phi-def}) [i.e., adding 
the linear equation (\ref{Phi-cond})] 
with subsequent 
calibration of the pulse 
(see Section~\ref{CAL}) 
will suppress the infidelity 
substantially toward or 
below $\lesssim 10^{-4}$.


\section{Author Contributions} 
All authors participated in 
the framing and discussion of the project 
and the evaluation of the 
results. R.B. performed all 
analytical and numerical 
calculations. 
All authors participated in 
the writing of the paper. 

\section{Data Availability}
Data and codes underlying this work 
are available from the corrsponding 
author upon reasonable request. 

\section{Conflicts of Interest}
The authors declare no conflicts of 
interest. 


\section{APPENDIX~A: 
Gate Simulator Matrix Elements}
\label{GS-MAT-EL}
In this section we present the matrix elements 
of the full Hamiltonian $\hat H_{MS}$ as well 
as the ones of the model 
Hamiltonians $\hat H^{(N_c,N_s)}$. 
We start with the matrix elements of 
the full Hamiltonian, $\hat H_{MS}$.
 
The matrix elements of the cosine part 
of $\hat H_{MS}$ are 
\begin{align}
C_{n_1 n_2\ldots  ; m_1m_2\ldots}^{(j)} &= 
\langle n_1 n_2 \ldots |  
\cos\left[ 
\sum_p \eta_p^j (\hat a_p^{\dagger} 
+\hat a_p) 
\right] 
|m_1 m_2\ldots\rangle 
\nonumber\\
&= 
\frac{1}{2} 
\langle n_1 n_2\ldots  |  
e^{i
\sum_p \eta_p^j (\hat a_p^{\dagger} 
+\hat a_p) } + 
e^{-i
\sum_p \eta_p^j (\hat a_p^{\dagger} 
+\hat a_p) }
|m_1 m_2\ldots\rangle 
\nonumber\\
&= 
\frac{1}{2} 
\left\{
\prod_p 
\langle n_p |  
e^{i \eta_p^j (\hat a_p^{\dagger} 
+\hat a_p) }
|m_p\rangle 
 + 
 \prod_p 
 \langle n_p | 
e^{-i \eta_p^j (\hat a_p^{\dagger} 
+\hat a_p) }
|m_p\rangle 
\right\} . 
\end{align}
We see that 
$C_{n_1 n_2\ldots ; m_1m_2\ldots}^{(j)}$
is real and symmetric in 
$n_p\leftrightarrow m_p$. 
Thus, we can arrange for 
$n_p\geq m_p$ for all $p$. 
With $n_p \geq m_p$ and 
\begin{equation}
\langle n|\, e^{\lambda\hat a^{\dagger}}\, 
|m\rangle = 
\begin{cases} 
0    &{\rm if\ } m > n\\
\frac{\lambda^{n-m}}
{(n-m)!}\, 
\left[\frac{n!}{m!}\right]^{1/2} 
&{\rm if\ } m\leq n .
\end{cases}
\label{elha-dagger}
\end{equation}
we have 
%
\begin{align}
C_{n_1 n_2\ldots ; m_1m_2\ldots}^{(j)} &= 
\frac{1}{2} 
\prod_p 
e^{-(\eta_p^j)^2/2} 
\left( \frac{m_p !}{n_p !}\right)^{1/2} 
\left(\eta_p^j\right)^{n_p-m_p} 
L_{m_p}^{(n_p-m_p)} 
\left[ (\eta_p^j)^2\right] 
\nonumber\\
&\left\{ 
\prod_p i^{(n_p-m_p)} + 
\prod_p (-i)^{(n_p-m_p)} 
\right\} . 
\end{align}
Now, define 
\begin{equation}
\sigma = \left[ 
\sum_p(n_p-m_p)\right] \ {\rm mod}\ 4. 
\end{equation}
Then: 
\begin{equation} 
C_{n_1 n_2\ldots  ; m_1m_2\ldots}^{(j)} = 
\prod_p 
e^{-(\eta_p^j)^2/2} 
\left( \frac{m_p !}{n_p !}\right)^{1/2} 
\left(\eta_p^j\right)^{n_p-m_p} 
L_{m_p}^{(n_p-m_p)} 
\left[ (\eta_p^j)^2\right] 
\begin{cases} 
1,      &{\rm if\ } \sigma = 0, \\ 
0,      &{\rm if\ } \sigma = 1, \\ 
-1,    &{\rm if\ } \sigma = 2, \\ 
0,     &{\rm if\ } \sigma = 3. 
\end{cases} 
\end{equation}
Because of 
\begin{equation}
C_{n_1 n_2\ldots ; m_1m_2\ldots}^{(j)} 
\sim 
\prod_p \left(\eta_p^j\right)^{n_p-m_p}, 
\end{equation}
and $|\eta_p^j|\ll 1$, 
we see that 
$C_{n_1 n_2\ldots  ; m_1m_2\ldots}^{(j)} $ 
is very close to 
diagonal, i.e., 
only the first few off-diagonals are 
significantly 
different from zero. 
This fact can be used to 
speed up the numerical 
integration of the system of 
linear equations significantly. 
 
Similarly, we obtain 
\begin{align}
&S_{n_1 n_2\ldots ; m_1m_2\ldots}^{(j)} = 
\langle n_1 n_2\ldots  |  
\sin\left[ 
\sum_p \eta_p^j (\hat a_p^{\dagger} 
+\hat a_p) 
\right] 
|m_1 m_2\ldots\rangle 
\nonumber\\
&= 
\prod_p 
e^{-(\eta_p^j)^2/2} 
\left( \frac{m_p !}{n_p !}\right)^{1/2} 
\left(\eta_p^j\right)^{n_p-m_p} 
L_{m_p}^{(n_p-m_p)} 
\left[ (\eta_p^j)^2\right] 
\begin{cases} 
0,      &{\rm if\ } \sigma = 0, \\ 
1,      &{\rm if\ } \sigma = 1, \\ 
0,    &{\rm if\ } \sigma = 2, \\ 
-1,     &{\rm if\ } \sigma = 3. 
\end{cases} 
\end{align}
%


%
\section{APPENDIX~B: THIRD-ORDER COMMUTATORS}
\label{3COMM}
In this appendix we list the results of 
all third-order commutators 
$T_{\alpha\beta\gamma}$ 
as defined in (\ref{T-def}) 
with $\eta$ orders $\eta^m$, $m\leq 4$. 
We group the commutators according 
to their $\eta$ order $m$.

Commutators $\hat T_{\alpha\beta\gamma}\sim \eta^0$:
%
\begin{equation}
\hat T_{000} = 0 ,
\label{T000}
\end{equation}

Commutators $\hat T_{\alpha\beta\gamma}\sim \eta^1$:
%
\begin{equation}
\hat T_{001} = \left(
\frac{2}{3}\right)
\sum_{j=1,2} \left[ 
\int_0^{\tau} g(t) G^2(t) 
\hat V_j(t)\, dt
\right] \hat\sigma_x^{(j)}, 
\label{T001} 
\end{equation}
\begin{equation}
\hat T_{010} = 2 \hat T_{001}, 
\label{T010}
\end{equation}
\begin{equation}
\hat T_{100} = 0, 
\label{T100}
\end{equation}

Commutators $\hat T_{\alpha\beta\gamma}\sim \eta^2$:
%
\begin{equation}
\hat T_{002} = 0, 
\label{T002}
\end{equation}
\begin{equation}
\hat T_{011} = \left(
\frac{8}{3}
\right)
\Phi[\eta,g] \ 
[\hat\sigma_x^{(1)}\hat\sigma_z^{(2)}
+\hat\sigma_x^{(2)}\hat\sigma_z^{(1)}],
\label{T011}
\end{equation}
\begin{equation}
\hat T_{020} = 0, 
\label{T0202}
\end{equation}
\begin{align}
\hat T_{101} &= 
\left(
-\frac{2}{3}
\right)\, 
\sum_{j=1,2}
\hat \sigma_y^{(j)} 
\int_0^{\tau} dt_1 
\int_0^{t_1} dt_2 
g(t_1) g(t_2) 
 \nonumber\\
&[G(t_1) - G(t_2) ]\,
[\hat V_j(t_1)\hat V_j(t_2) +
\hat V_j(t_2)\hat V_j(t_1)], 
\label{T101}
\end{align}
\begin{align}
\hat T_{110} &= \left(
-\frac{2}{3}
\right)
\int_0^{\tau} dt_1 
\int_0^{t_1} dt_2
g(t_1) g(t_2) G(t_1)\,  
\nonumber\\
&\sum_j 
[\hat V_j(t_1)\hat V_j(t_2)+
\hat V_j(t_2)\hat V_j(t_1)]
\hat\sigma_y^{(j)}
\nonumber\\
&+ \left(
\frac{4}{3}
\right) \Phi[\eta,g]\ 
[\hat\sigma_x^{(1)}\hat\sigma_z^{(2)}
+\hat\sigma_x^{(2)}\hat\sigma_z^{(1)}], 
\label{T110}
\end{align}
\begin{equation}
\hat T_{200} = 0 . 
\label{T200}
\end{equation}

Commutators $\hat T_{\alpha\beta\gamma}\sim \eta^3$:
%
\begin{equation}
\hat T_{003} = 
\left(-
\frac{1}{9}
\right)
\sum_j 
\left\{
\int_0^{\tau} 
g(t) G^2(t) \hat V_j^3(t)\, dt 
\right\}
\, \hat\sigma_x^{(j)},
\label{T003}
\end{equation}
\begin{align}
\hat T_{012} &= 
\left(\frac{1}{6}\right)
\sum_k \hat\sigma_x^{(k)} 
\int_0^{\tau} dt_1 
\int_0^{t_1} dt_2 g(t_1) g(t_2) 
\nonumber\\
&\Big\{ 
G(t_2)[\hat V_k(t_2)\hat V_k^2(t_1)
+\hat V_k^2(t_1) \hat V_k(t_2)]
\nonumber\\
&-G(t_1) 
[\hat V_k(t_1)\hat V_k^2(t_2)
+\hat V_k^2(t_2) \hat V_k(t_1)]
\Big\} 
\nonumber\\
&- 
\left(\frac{2}{3}\right) 
\sum_p \eta_p^1 \eta_p^2 
\int_0^{\tau} dt_1 
\int_0^{t_1} dt_2 g(t_1) g(t_2) 
\sin[\omega_p(t_1-t_2)] 
\nonumber\\
&\Big\{ 
[G(t_2)\hat V_1(t_1) + G(t_1)\hat V_1(t_2)] 
\hat\sigma_y^{(1)}\hat\sigma_z^{(2)} 
\nonumber\\
&+[G(t_2)\hat V_2(t_1)+G(t_1)\hat V_2(t_2)]
\hat\sigma_y^{(2)}\hat\sigma_z^{(1)} 
\Big\} , 
\label{T012}
\end{align}
\begin{align}
\hat T_{021} &= 
\left(\frac{1}{3}\right)
\sum_k \hat\sigma_x^{(k)} 
\int_0^{\tau} dt_1 
\int_0^{t_1} dt_2 g(t_1) g(t_2) 
G(t_1)
\nonumber\\
&[\hat V_k^2(t_1)\hat V_k(t_2)
+\hat V_k(t_2) \hat V_k^2(t_1)]
\nonumber\\
&- 
\left(\frac{2}{3}\right) 
\sum_p \eta_p^1 \eta_p^2 
\int_0^{\tau} dt_1 
\int_0^{t_1} dt_2 g(t_1) g(t_2) 
\sin[\omega_p(t_1-t_2)] 
\nonumber\\
&\Big\{ 
[G(t_2)\hat V_1(t_2) + G(t_1)\hat V_1(t_1)] 
\hat\sigma_y^{(1)}\hat\sigma_z^{(2)} 
\nonumber\\
&+[G(t_2)\hat V_2(t_2)+G(t_1)\hat V_2(t_1)]
\hat\sigma_y^{(2)}\hat\sigma_z^{(1)} 
\Big\} , 
\label{T021}
\end{align}
\begin{equation}
\hat T_{030} = 
 \left(
-\frac{2}{9}
\right)
\sum_{j=1,2}
\hat\sigma_x^{(j)}
\int_0^{\tau} 
g(t) G^2(t) \hat V_j^3(t)\ dt ,
\label{T030}
\end{equation}
 \begin{equation}
 \hat T_{102} = 0, 
 \label{T102}
 \end{equation}
\begin{equation}
\hat T_{111} = 0,
\label{T111}
\end{equation}
\begin{equation}
\hat T_{120} = 0, 
\label{T120}
\end{equation}
\begin{align}
\hat T_{201} &= 
\left(-\frac{1}{6}\right) 
\sum_j \hat\sigma_x^{(j)} 
\int_0^{\tau} dt_1\int_0^{t_1} dt_2 
g(t_1) g(t_2) 
[G(t_1)-G(t_2)] 
\nonumber\\
&\left\{
\hat V_j^2(t_1)\hat V_j(t_2) + 
\hat V_j(t_2)\hat V_j^2(t_1) +
\hat V_j^2(t_2)\hat V_j(t_1) + 
\hat V_j(t_1)\hat V_j^2(t_2)
\right\}
\nonumber\\
&+ 
\left(\frac{2}{3}\right) 
\sum_{p,j\neq k} \hat\sigma_y^{(j)} 
\hat\sigma_z^{(k)} 
\eta_p^j \eta_p^k 
\nonumber\\
&\int_0^{\tau} dt_1\int_0^{t_1} dt_2 
g(t_1) g(t_2) 
[G(t_1)-G(t_2)] 
\sin[\omega_p(t_1-t_2)] 
\, 
[\hat V_j(t_1)-\hat V_j(t_2)] , 
\label{T201}
\end{align}
\begin{align}
\hat T_{210} &= 
\left(\frac{1}{6}\right) 
\sum_j \hat\sigma_x^{(j)} 
\int_0^{\tau} dt_1\int_0^{t_1} dt_2 
g(t_1) g(t_2)  
\nonumber\\
&\left\{
G(t_2)[\hat V_j^2(t_1)\hat V_j(t_2) + 
\hat V_j(t_2)\hat V_j^2(t_1)] -
G(t_1)[\hat V_j^2(t_2)\hat V_j(t_1) + 
\hat V_j(t_1)\hat V_j^2(t_2)]
\right\}
\nonumber\\
&-
\left(\frac{2}{3}\right) 
\sum_{p,j\neq k} \hat\sigma_y^{(j)} 
\hat\sigma_z^{(k)} 
\eta_p^j \eta_p^k 
\int_0^{\tau} dt_1\int_0^{t_1} dt_2 
g(t_1) g(t_2)  
\sin[\omega_p(t_1-t_2)] 
\nonumber\\
&[G(t_2)\hat V_j(t_1)+G(t_1)\hat V_j(t_2)] , 
\label{T210}
\end{align}
\begin{equation}
\hat T_{300} = 0.
\label{T300}
\end{equation}

Commutators $\hat T_{\alpha\beta\gamma}\sim \eta^4$:
%
%
\begin{equation}
\hat T_{004} = 0,
\label{T004}
\end{equation}
\begin{align}
\hat T_{013} &= 
\left(
-\frac{1}{3}
\right)\,
[\hat\sigma_x^{(1)}\hat\sigma_z^{(2)}
+\hat\sigma_x^{(2)}\hat\sigma_z^{(1)}]
\nonumber\\
&\sum_{p,j=1,2} \eta_p^1\eta_p^2 
\int_0^{\tau} dt_1 
\int_0^{t_1} dt_2 
g(t_1) g(t_2) 
\sin[\omega_p(t_1-t_2)]
\nonumber\\
&\ \ \ [G(t_1) \hat V_j^2(t_2) + 
G(t_2) \hat V_j^2(t_1)], 
\label{T013}
\end{align}
 \begin{equation}
 \hat T_{022} = 0, 
 \label{T022}
 \end{equation}
\begin{align}
\hat T_{031} &= 
\left(
-\frac{2}{3}
\right)\,
[\hat\sigma_x^{(1)}\hat\sigma_z^{(2)}
+\hat\sigma_x^{(2)}\hat\sigma_z^{(1)}]
\nonumber\\
&\sum_{p,j=1,2} \eta_p^1\eta_p^2 
\int_0^{\tau} dt_1 
\int_0^{t_1} dt_2 
g(t_1) g(t_2) G(t_1) 
\sin[\omega_p(t_1-t_2)]
\hat V_j^2(t_1),
\label{031}
\end{align}
\begin{equation}
\hat T_{040} = 0, 
\label{T040}
\end{equation}
\begin{align}
\hat T_{103} &= 
\int_0^{\tau} dt_1 
\int_0^{t_1} dt_2 
g(t_1) g(t_2) [G(t_1) - G(t_2) ]
\nonumber\\
&\Bigg\{ 
\left(
\frac{1}{18}
\right)\,
\sum_{j=1,2} \hat\sigma_y^{(j)} 
[\hat V_j(t_1)\hat V_j^3(t_2) + 
\hat V_j^3(t_2)\hat V_j(t_1) +
\hat V_j(t_2)\hat V_j^3(t_1) +
\hat V_j^3(t_1)\hat V_j(t_2)] 
\nonumber\\
&+
\left(
\frac{1}{3}
\right)\,
\sum_{j\neq k} 
\hat\sigma_x^{(j)}\hat\sigma_z^{(k)}
\sum_p \eta_p^j \eta_p^k
[\hat V_k^2(t_2)-\hat V_k^2(t_1)]
\sin[\omega_p(t_1-t_2)]
\Bigg\},
\label{T103}
\end{align}
\begin{align}
\hat T_{112} &= 
\left(\frac{1}{6}\right)
\int_0^{\tau} dt_1 
\int_0^{t_1} dt_2 
\int_0^{t_2} dt_3 
g(t_1) g(t_2) g(t_3) 
\nonumber\\
&\Bigg\{ 
\left(-\frac{1}{2}\right) 
\sum_j \hat\sigma_y^{(j)} 
\Big\{ 
\hat V_j(t_1)[\hat V_j(t_2) 
\hat V_j^2(t_3) + 
\hat V_j^2(t_3) \hat V_j(t_2)] 
\nonumber\\
&+ 
[\hat V_j(t_2)\hat V_j^2(t_3) + 
\hat V_j^2(t_3) \hat V_j(t_2)]
\hat V_j(t_1)
+
\hat V_j(t_3)[\hat V_j(t_2) 
\hat V_j^2(t_1) + 
\hat V_j^2(t_1) \hat V_j(t_2)] 
\nonumber\\
&+ 
[\hat V_j(t_2)\hat V_j^2(t_1) + 
\hat V_j^2(t_1) \hat V_j(t_2)]
\hat V_j(t_3) \Big\} 
\nonumber\\
&-2\sum_{p,j\neq k} 
\hat\sigma_x^{(j)}\hat\sigma_z^{(k)}
\eta_p^j \eta_p^k 
\nonumber\\
&\Big\{ 
\hat V_k^2(t_3)\sin[\omega_p(t_1-t_2)] 
- V_k^2(t_1) \sin[\omega_p(t_2-t_3)] 
\nonumber\\
&+[\hat V_k(t_2)\hat V_k(t_3) 
+\hat V_k(t_3)\hat V_k(t_2)] 
\sin[\omega_p(t_1-t_3)]
\nonumber\\
&-[\hat V_k(t_2)\hat V_k(t_1) 
+ \hat V_k(t_1)\hat V_k(t_2)]
\sin[\omega_p[t_1-t_3)] 
\Big\} 
\nonumber\\
&+4\sum_{pq} 
\eta_p^1\eta_p^2\eta_q^1\eta_q^2
(\hat\sigma_y^{(1)}+\hat\sigma_y^{(2)}) 
\sin[\omega_q(t_1-t_3)]
\nonumber\\
&\left\{
\sin[\omega_p(t_2-t_3)]
-\sin[\omega_p(t_2-t_1)]
\right\}
\nonumber\\
&- 
2\sum_p
\eta_p^1\eta_p^2
\hat\sigma_x^{(1)}\hat\sigma_z^{(2)}
[\hat V_2(t_1)\hat V_2(t_3) + 
\hat V_2(t_3)\hat V_2(t_1)] 
\nonumber\\
&\left\{
\sin[\omega_p(t_2-t_3)]
+\sin[\omega_p(t_2-t_1)]
\right\}
\nonumber\\
&- 
2\sum_p
\eta_p^1\eta_p^2
\hat\sigma_x^{(2)}\hat\sigma_z^{(1)}
[\hat V_1(t_1)\hat V_1(t_3) + 
\hat V_1(t_3)\hat V_1(t_1)] 
\nonumber\\
&\left\{
\sin[\omega_p(t_2-t_3)]
+\sin[\omega_p(t_2-t_1)]
\right\}
\Bigg\} , 
\label{T112}
\end{align}
\begin{align}
\hat T_{121} &= 
\left(-\frac{1}{12}\right)
\int_0^{\tau} dt_1 
\int_0^{t_1} dt_2
\int_0^{t_2} dt_3 
g(t_1) g(t_2) g(t_3) 
\nonumber\\ 
&\sum_{jkl} 
\Big\{ 
2[\hat V_j(t_1)\hat V_k^2(t_2)\hat V_l(t_3) 
+\hat V_j(t_3)\hat V_k^2(t_2)\hat V_l(t_1)]
\hat\sigma_x^{(j)}\hat\sigma_y^{(k)}
\hat\sigma_x^{(l)} 
\nonumber\\
&-
[\hat V_j(t_1)\hat V_k(t_3)\hat V_l^2(t_2)
+\hat V_j(t_3)\hat V_k(t_1)\hat V_l^2(t_2)]
\hat\sigma_x^{(j)}\hat\sigma_x^{(k)}
\hat\sigma_y^{(l)}
\nonumber\\
&-
[\hat V_j^2(t_2)\hat V_k(t_3)\hat V_l(t_1)
+\hat V_j^2(t_2)\hat V_k(t_1)\hat V_l(t_3)]
\hat\sigma_y^{(j)}\hat\sigma_x^{(k)}
\hat\sigma_x^{(l)}
\Big\} , 
\label{T121}
\end{align}
\begin{align}
\hat T_{130} &= 
\left(
\frac{1}{9}
\right)\, 
\int_0^{\tau} dt_1 
\int_0^{t_1} dt_2 
g(t_1) g(t_2) G(t_1)
\nonumber\\
&\left\{\sum_{j=1,2}\sigma_y^{(j)} 
[\hat V_j(t_2) \hat V_j^3(t_1) + 
\hat V_j^3(t_1) \hat V_j(t_2)]
\right\}
\nonumber\\
&-
\left(
\frac{2}{3}
\right)\,
\int_0^{\tau} dt_1 
\int_0^{t_1} dt_2 
g(t_1) g(t_2) 
\nonumber\\
&\sum_{\substack{j,k=1,2 \\ j\neq k}}
\sum_{p=1}^N
\eta_p^j\eta_p^k
\sigma_x^j\sigma_z^k
G(t_1)\hat V_k^2(t_1)
\sin[\omega_p(t_1-t_2)],
\label{T130}
\end{align}
\begin{equation}
\hat T_{202} = 0
\label{T202}
\end{equation}
\begin{align}
 \hat T_{211} &= 
 \left(
 \frac{4}{3} 
  \right)
 \sum_{p,k\neq j} 
 \eta_p^1 \eta_p^2 
 \hat \sigma_z^{(j)}
 \hat \sigma_x^{(k)}
\nonumber\\
&\int_0^{\tau} dt_1 
\int_0^{t_1} dt_2
\int_0^{t_2} dt_3
g(t_1) g(t_2) g(t_3) 
\hat V_j^2(t_1) 
\sin[\omega_p(t_2-t_3)] , 
\label{T211}
\end{align}
\begin{equation}
\hat T_{220} = 0 , 
\label{T220}
\end{equation}
\begin{align}
\hat T_{301} &= 
\left(
\frac{1}{18}
\right)\,
\int_0^{\tau} dt_1 
\int_0^{t_1} dt_2 
g(t_1) g(t_2) [G(t_1) - G(t_2) ]
\nonumber\\
&\left\{\sum_{j=1,2} 
\sigma_y^{(j)} 
[ V_j^3(t_1) V_j(t_2) + 
   V_j(t_2) V_j^3(t_1) + 
   V_j^3(t_2) V_j(t_1) + 
   V_j(t_1) V_j^3(t_2)
  ] \right\}
\nonumber \\
&+
\left(
\frac{1}{3}
\right)\,
\int_0^{\tau} dt_1 
\int_0^{t_1} dt_2 
g(t_1) g(t_2) [G(t_1) - G(t_2) ]
\nonumber\\
&\sum_{\substack{j,k=1,2 \\ j\neq k}}
\sum_{p=1}^N
\eta_p^j\eta_p^k
\sigma_x^j\sigma_z^k
[\hat V_j^2(t_1) -
\hat V_j^2(t_2)]
\sin[\omega_p(t_1-t_2)],
\label{T301}
\end{align}
\begin{align}
\hat T_{310} &= 
\left(
\frac{1}{18}
\right)\,
\int_0^{\tau} dt_1 
\int_0^{t_1} dt_2 
g(t_1) g(t_2) 
\nonumber\\
&\Bigg[
\sum_{j=1,2}\hat\sigma_y^{(j)} 
\Big\{ G(t_1) 
[\hat V_j^3(t_2)\hat V_j(t_1) + 
 \hat V_j(t_1) \hat V_j^3(t_2)]
 \nonumber\\
 &- G(t_2) 
[\hat V_j^3(t_1)\hat V_j(t_2) + 
 \hat V_j(t_2) \hat V_j^3(t_1)]
\Big\} 
\Bigg] 
\nonumber\\
&-\left(
\frac{1}{3}
\right)\,
\int_0^{\tau} dt_1 
\int_0^{t_1} dt_2 
g(t_1) g(t_2) \sin[\omega_p(t_1-t_2)]
\nonumber\\
&\sum_{\substack{j,k=1,2 \\ j\neq k}}
\sum_{p=1}^N
\eta_p^j\eta_p^k
\sigma_x^j\sigma_z^k
[G(t_1)\hat V_j^2(t_2) +
G(t_2)\hat V_j^2(t_1)].
\label{T310}
\end{align}
\begin{equation}
\hat T_{400} = 0 . 
\label{T400}
\end{equation}
%
%


%
\bibliographystyle{apsrev4-1}
\bibliography{Q-comp}
\end{document}